\documentclass{ametsocV6.1}

\usepackage{amsmath}
\usepackage{soul}
\usepackage{enumitem}
\usepackage{booktabs}

\setlist[enumerate]{itemsep=0mm}

\newcommand{\Vf}{V_a}

\title{A Deep Learning-based Velocity Dealiasing Algorithm Derived from the WSR-88D Open Radar Product Generator}

\authors{Mark S. Veillette,\aff{a}\correspondingauthor{Mark S. Veillette, mark.veillette@ll.mit.edu} 
James M. Kurdzo,\aff{a} 
Phillip M. Stepanian,\aff{a} 
Joseph McDonald,\aff{a} 
Siddharth Samsi,\aff{a} 
 and John Y. N. Cho\aff{a} 
}

\affiliation{\aff{a}{Lincoln Laboratory, Massachusetts Institute of Technology, Lexington, Massachusetts}\\
}

\abstract{ Radial velocity estimates provided by Doppler weather radar are critical measurements used by operational forecasters for the detection and monitoring of life-impacting storms.   The sampling methods used to produce these measurements are inherently susceptible to aliasing, which produces ambiguous velocity values in regions with high winds, and needs to be corrected using a velocity dealiasing algorithm (VDA).  In the US, the Weather Surveillance Radar -- 1988 Doppler (WSR-88D) Open Radar Product Generator (ORPG) is a processing environment that provides a world-class VDA; however, this algorithm is complex and can be difficult to port to other radar systems outside of the WSR-88D network.  In this work, a Deep Neural Network (DNN) is used to emulate the 2-dimensional WSR-88D ORPG dealiasing algorithm.  It is shown that a DNN, specifically a customized U-Net, is highly effective for building VDAs that are accurate, fast, and portable to multiple radar types.  To train the DNN model, a large dataset is generated containing aligned samples of folded and dealiased velocity pairs.  This dataset contains samples collected from WSR-88D Level-II and Level-III archives, and uses the ORPG dealiasing algorithm output as a source of truth.  Using this dataset, a U-Net is trained to produce the number of folds at each point of a velocity image.   Several performance metrics are presented using WSR-88D data.  The algorithm is also applied to other non-WSR-88D radar systems to demonstrate portability to other hardware/software interfaces.  A discussion of the broad applicability of this method is presented, including how other Level-III algorithms may benefit from this approach.

} 

\begin{document}

\maketitle

%
%
%
\statement
	 Accurate and timely estimates of wind within storms are critically important for a number of applications, including severe storm nowcasting, maritime operational planning, aviation forecasting, and public safety coordination.  Velocity aliasing is a common artifact that requires data quality control.  While velocity dealiasing algorithms (VDAs) have been  developed for decades, they remain a computationally complex and challenging problem.  This paper presents an application of deep neural networks (DNNs) to increase the computational efficiency and portability of VDAs.  A DNN is trained to emulate an operational algorithm and performance is quantified over a large dataset.  This work gives a convincing example of the benefits that deep learning can provide for radar algorithms, and future work  highlighting these opportunities is discussed.

%
%

%

\section{Introduction}\label{sec:intro}


Meteorological radars are an integral piece of the weather enterprise's toolset for observing and predicting weather with relatively high spatial and temporal resolution and across broad swaths of area compared with in situ observation methods.  Weather radar observations are essential for a wide range of meteorological applications including quantitative precipitation estimation \citep[QPE;][]{hong+15}, tornado detection and warning \citep[e.g.,][]{brown++78}, identifying severe wind and hail hazards \citep[e.g.,][]{burgess+90,ortega++16}, characterizing tornado/mesoscale dynamics \citep[e.g.,][]{wurman++96}, forecasting real-time mesoscale snow bands \citep[e.g.,][]{kristovich++03} and hurricane rain bands \citep[e.g.,][]{biggerstaff++21}, and assimilation into numerical weather prediction systems \citep[e.g.,][]{stensrud++09}. 

To enable these applications, the United States (US) National Weather Service (NWS) operates a network of Weather Surveillance Radar -- 1988 Doppler (WSR-88D) meteorological radars that provide real-time atmospheric observations over the US and outlying territories. The ability of WSR-88D estimates to successfully support the NWS mission is contingent on a series of processing steps that convert raw in-phase and quadrature receiver base-band samples (I/Q; termed ``Level-I'' data) into a collection of standard radar data. The most basic of these data for a dual-polarization radar are the six ``Level-II'' variables that are calculated through time-series processing of the Level-I signals: horizontal reflectivity factor, radial velocity, Doppler spectrum width, differential reflectivity, co-polar correlation coefficient, and differential phase. These Level-II variables contain information relating to the size, shape, abundance, composition, and motion of objects in the atmosphere. A number of additional ``Level-III'' products are generated within the WSR-88D Open Radar Product Generator (ORPG) using more-complex workflows, and often incorporate 2-, 3-, or 4-dimensional analyses across multiple Level-II variables, sometimes in combination with ancillary datasets beyond just radar (e.g., model data). For example, the ORPG QPE algorithm combines multiple Level-II variables and integrates estimated rainfall over time in a Level-III product.  As a result of these sophisticated processing techniques, Level-III products often have a greater degree of quality control, are more directly relatable to common users of radar data, and provide more intuitive visual interpretation of the phenomena they are meant to characterize, making them well suited for use by weather forecasters and the general public.

\begin{figure}[ht]
  \centerline{\includegraphics[width=36pc]{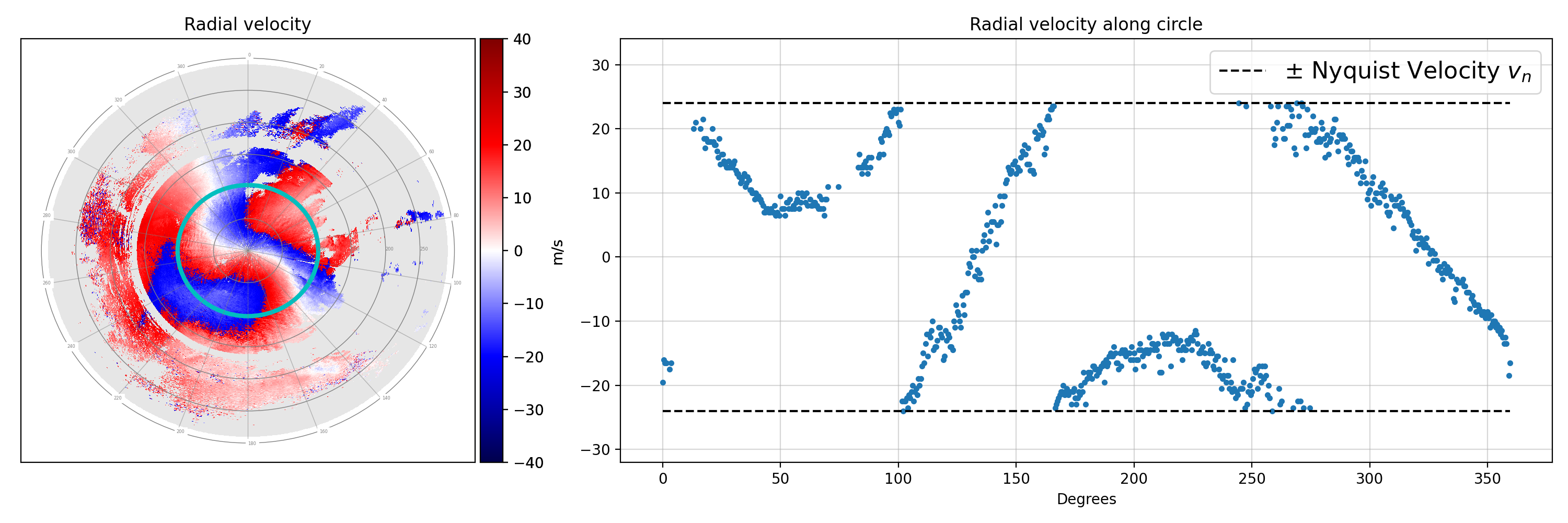}}
   \caption{An example of velocity aliasing, shown as a PPI of a velocity field (left) and the raw velocity values when traversing the green circle (right).  In the PPI, the areas of outbound (inbound) radial velocities toward the southwest (northeast) ``flip'' to the opposite sign when the Nyquist values of +/-$V_n=24$ m s$^{-1}$ are exceeded.  In the plot on the right, the green circle from the left image is traversed, showing discontinuities that ``jump'' from the positive Nyquist value to the negative Nyquist value and vice versa.  The radial velocity color table for the PPI here and throughout is courtesy \citet{collis++18}.}\label{fig:da_example}
 \end{figure}

One of the major challenges in weather radar is data quality control.  In particular, range-Doppler ambiguities, a direct result of the Doppler dilemma \citep{doviak+93,skolnik02}, are a significant challenge for Doppler weather radars.  The main focus of this work is velocity aliasing, which causes ambiguous velocity values to be estimated in regions where winds exceed the Nyquist velocity, $V_n$. The result of this ambiguity is that the magnitude of the velocity vector---the wind speed relative to the radar---can be any of several different values, and even the sign of the radial wind direction---either toward or away from the radar---is unknown. Applications relying on accurate wind measurements require eliminating these ambiguities, revealing the singular `true' radial wind velocity value. Any winds that exceed a nominal maximum unambiguous velocity are subject to aliasing. This maximum  unambiguous velocity is a function of the radar wavelength and the pulse repetition frequency (PRF; see Eq. \ref{eq:nyquist}).  Too high of a PRF results in unambiguous ranges that are too short for operational use.  At higher frequencies, such as C-band and X-band radars, the Doppler dilemma becomes even more challenging due to the shorter wavelengths.  This relation is given by \citep{doviak+93}:

\begin{equation}
    V_n = \frac{\lambda \ \mathrm{PRF}}{4},
    \label{eq:nyquist}
\end{equation}

\noindent where the Nyquist velocity $V_n$ is in m s$^{-1}$, $\lambda$ is the wavelength in m, and PRF is the pulse repetition frequency in s$^{-1}$.  An example of velocity aliasing is shown in Fig. \ref{fig:da_example}, where a plan position indicator (PPI) of radial velocity in the left panel alternates between inbounds (negative values/blue colors) and outbounds (positive values/red colors) when traversing along the green circle.  In the right panel, the aliasing effect of the velocities is observed as discontinuities that jump across the Nyquist intervals.  Given the tradeoff of the PRF selection, velocity dealiasing algorithms (VDAs) are important for any meteorological radar.  Velocity dealiasing is particularly important in cases of severe weather and hurricanes where radial velocities are high.  In tornadoes, for example, Doppler velocities can easily exceed the Nyquist velocity several times \citep[e.g.,][]{burgess++02}.  Level-III algorithms such as the tornadic vortex signature \citep[TVS;][]{brown++78,mitchell++98} and mesocyclone detection algorithm \citep[MDA;][]{stumpf++98} depend on accurate velocity dealiasing.

While the WSR-88D VDA demonstrates exceptional performance and ongoing improvements within each ORPG Build, the algorithm is relatively complex and requires close to 1 s per tilt (constant-elevation-angle 360$^{\circ}$ azimuthal scan) for complex storm environments.  While this is acceptable on appropriate hardware and for the current scan rates for the WSR-88D (which peak at 4.8 rpm and must keep up in ``real time''), there are several reasons why a simpler/faster implementation would be highly desirable.  For example, future radar systems, such as phased arrays, may scan much faster \citep{weber++21}, resulting in the need for algorithms that are capable of running more efficiently.  Several ORPG algorithms take much more than 1 s to run, making more-efficient performance for future faster-scanning radars a must.  In addition, the ORPG VDA utilizes environmental wind information, which is not always accessible in real-time or when processing large batches of archived data.  An accurate, fast, and highly portable algorithm that only requires a single velocity image would be of great value in situations like these.

For these reasons, we utilize Machine Learning (ML), specifically Deep Learning (DL), to emulate the 2-dimensional WSR-88D VDA  using purely data-driven approaches.  DL approaches have been successfully demonstrated in diverse meteorological applications such as precipitation nowcasting \citep{ravuri2021skilful,xingjian2015convolutional},  sub-seasonal forecasting \citep{weyn2021sub},  satellite precipitation estimation, \citep{wang2021precipgan}, synthetic weather radar~\citep{veillette++18,hilburn2021development}, mesocyclone classification \citep{Krinitskiy2018DeepCN}, extreme weather event detection \citep{Racah2017ExtremeWeatherAL}, and several others.  Within radar meteorology specifically, a range of different ML algorithms have been used, and many were summarized by \citet{chandra20}, with foci on QPE, precipitation classification, and nowcasting.  For example, \citet{chen++19} use deep neural networks to estimate precipitation using both satellite-borne and ground-based radar networks.  Other QPE-based DL approaches include those by \citet{yo++21}, \citet{husnoo++21} (using the U.K. weather radar network), and \citet{peng++21}.  Although not DL-centric, precipitation classification has used other forms of ML, including fuzzy logic \citep{park++09} and Bayesian approaches \citep{yang+19}.

Currently, only one operational ORPG algorithm utilizes modern machine learning techniques \citep[the chaff detection algorithm;][]{kurdzo++17b}, but due to the wide accessibility to years of archived data, the resources exist to enable other ML-based algorithms \citep[e.g.,][]{jatau++21}.  A DL-based VDA should enable a faster algorithm, provide a WSR-88D-quality dealiaser for users of other radar datasets, and possibly be implementable within open source software packages.  The VDA is an ideal algorithm to attempt to replicate within the ORPG, as it uses just one moment (the velocity field), is computationally complex (with a desire to decrease resources, especially for non-ORPG hardware), is a critical need for non-WSR-88D radars, and performs well enough to use as an unsupervised truth dataset.  In this work, we specifically attempt to  \textit{emulate} the WSR-88D velocity dealiasing algorithm.  This is an important distinction, as we are not expecting to necessarily improve upon the algorithm.  The reason for this approach is that the existing algorithm can serve as an exceptionally large truthed dataset.  The Level-III velocity product can be considered truth at all times if we are attempting to emulate/recreate the algorithm, eliminating the need for any manual dealiasing.

While improving upon the algorithm would be ideal, this approach sets the stage for future re-creation of other ORPG algorithms using ML techniques for both computational efficiency and portability.  We explore different techniques to optimize the algorithm, approaches for improving upon baseline results, and applications beyond the WSR-88D to other radars.  The paper is organized as follows: Section 2 discusses prior work in the VDA space; Section 3 describes the data and the approach; Section 4 discusses the chosen model architecture; Section 5 presents results; and Section 6 provides a discussion of the applicability of the results along with some conclusions.

\section{Prior Work}\label{ss:prior_work}

Since some of the earliest work on velocity ambiguities by \citet{ray+77}, several approaches to dealiasing constant-PRF radar radial velocity data have been presented in the literature, and a subset of these have been deployed operationally in the ORPG.  The initial ORPG VDA was based on \citet{eilts+90}, which takes a  radial-by-radial approach to unfolding velocities.  Each radial is compared with the previous radial using a local environment dealiasing (LED) technique.  These radial continuity checks are based only in the radial dimension, making the algorithm computationally efficient.  As discussed in \citet{witt++09}, as well as \citet{zittel++11}, the \citet{eilts+90} approach had worked well in an initial ORPG implementation with limited computational resources, but often failed in cases of lower Nyquists (i.e., volume coverage pattern 31) and strong shear (mesocyclones, gust fronts, etc.).  As new computational capabilities were added, it became possible to implement a new ORPG VDA based on the work in \citet{jing+93}, which takes a two-dimensional approach to solving a linear system of equations in a connected two-dimensional region by minimizing discontinuities in a least-squares problem.  The ORPG version of the algorithm adds a weighting factor to Eq. 5 of \citet{jing+93}, which aids with issues in noisy data \citep{witt++09}.  A global dealiaser using environmental data is also applied as part of \citet{jing+93}, and is used in the ORPG VDA as the second step of a ``two-phased dealiasing approach'' \citep{witt++09}.   Anecdotally, the ORPG VDA performs admirably in most situations, however, the two-dimensional VDA is still relatively complex and computationally expensive \citep{witt++09,zittel++11,code21}.

The current WSR-88D VDA is available via the ORPG Common Operations and Development Environment (CODE) software package \citep{code21}, an open-source version of the ORPG.  As of this writing, the CODE ORPG Build 20 source code (written mainly in \texttt{C}) is available to the public via download at \texttt{https://www.weather.gov/code88d/}.  To the authors' knowledge, no widespread use of the WSR-88D VDA is used in other operational radars or research radars.  

Beyond 2D VDAs (the main focus of this work), other approaches that utilize additional dimensions have been developed.  For example, \citet{james+01} present a real-time, four-dimensional dealiasing (4DD) scheme that takes a top-tilt-down approach using the elevation and time dimensions in addition to investigating neighboring gates and radials.  This technique is implemented along with other dealiasing algorithms in the Python ARM Radar Toolkit (\texttt{Py-ART}) \citep{helmus+16}.  These algorithms are also often similarly (or more) computationally complex compared with the WSR-88D algorithm, frequently taking a non-trivial amount of time to run, especially in complex velocity field situations.  Additionally, at C band and higher frequencies, with low Nyquist values, automated dealiasers perform particularly poorly, especially in tornadoes \citep[e.g.,][]{bluestein++07,yu++07a}.  For this reason, most research radars (e.g., mobile C-band, X-band, Ka-Band, and W-band radars) that collect tornado data must have their velocities \textit{manually} dealiased, step-by-step in time and elevation \citep[e.g.,][]{oye++95}.  This onerous task is extremely time consuming and often unavoidable.  Even operational radars at C band, especially those in areas of sparse data and high environmental shear, can suffer from dealiaser failures, such as the Australian weather radar network \citep{louf++20}.





After the 4DD approach was developed, additional dimensions were explored for dealiasing, especially with low-Nyquist radar systems.  For example, \citet{zhang+06} discuss multipass filters that determine confidence levels that subsequently increase with additional passes of dealiasing.  Similar to other multi-dimensional approaches, multiple passes add time and complexity to the processing steps despite performing exceptionally well.  \citet{xu++11} used VAD profiles to dealias radar data for numerical weather prediction assimilation.  While their methods perform well in many case examples, they tend to break down in areas of localized high shear (mesocyclones, tornadoes, hurricane eyewalls, etc.).  \citet{he++12} developed a C-band-specific approach to dealiasing for the Chinese CINRAD radar network, which operates at a single PRF with a high frequency, making them particularly susceptible to velocity aliasing issues.  

More recently, the Unfold Radar Velocity (UNRAVEL) algorithm takes a modular approach to a VDA, allowing for an optimization of results that can balance between performance needs and computational complexity \citep{louf++20}.  UNRAVEL currently contains two different (but expandable) dealiasing strategies: a tilt-by-tilt (2-dimensional) and a 3-dimensional continuity check.  \citet{louf++20} use artificially aliased S-band data as a reference for analyzing performance at C band, a technique that is adopted in a similar fashion in this study.  Finally, \citet{feldmann++20} introduce the region-based recursive Doppler dealiasing method (R2D2) for dealiasing in highly sheared environments, a historically difficult area for VDAs.  A 4-dimensional procedure with a 2-dimensional continuity check iterates until improvement is no longer achieved, and the highly sheared areas are given a ``focus'' with an image mask and a spatial buffer, resulting in exceptional performance.

\section{Data Description}

To establish notation, we will represent the true radial velocity $V$ at a gate as the sum of the observed velocity that has potentially been subject to aliasing, $\Vf$, and an even multiple of the Nyquist velocity $V_n$, i.e.,
\begin{equation}\label{e:actv}
    V = \Vf + 2 n V_n,
\end{equation}
where $n\in \mathbb{Z}$ represents the number of ``folds'' required to dealias the observed folded velocity.  We represent a single tilt of $\Vf$ and $V$ as a two-dimensional array of shape $N_{az} \times N_{rng}$ representing the azimuthal and radial dimensions, respectively.  The goal of a VDA is to infer the correct value of $n$ at each gate so that the true velocity can be reconstructed using Eq. \ref{e:actv}.  In this work, this task is accomplished by learning a deep neural network (DNN) $n = f(\Vf;\theta)$ that outputs the correct fold number for all gates using a single tilt of $\Vf$.  The variable $\theta$ in this function denotes weights of a DNN that will be learned during training.  Note that a 2-dimensional VDA was chosen to maximize portability of the algorithm by not requiring additional inputs such as environmental winds.  However, the methodologies described below can be extended to higher-dimensional versions of the problem (e.g., 4DD) that incorporate additional tilts, time steps, radar variables, environmental winds, and/or other inputs if desired.  This is an important aspect regarding the potential for DL to vastly improve the run-times of these more-complex algorithms.

This proposed VDA is entirely data-driven, and requires a large number of aligned samples containing input aliased velocity, associated Nyquist velocities, and dealiased target velocity: $(V_{a_{,i}},V_{n,i},V_i)$.  Below we describe the sources used to build datasets for the development and evaluation of a DNN-based VDA.  In addition, we describe the processing pipeline implemented to automatically construct labeled training samples required for DNN training and quantitative assessments.

\subsection{Training Data Description}

Training data for dealiasing was derived entirely from WSR-88D data products.  As described in the introduction, WSR-88D data is available from two tiers of products, Level-II variables  and Level-III products that are derived from Level-II variables.  Of particular interest for this work is the Level-III radial velocity product, which is a dealiased and data quality edited version of the (potentially aliased) Level-II velocity.

While no perfect ``truth'' exists for this task (outside manually labeled data), Level-III velocity was chosen as truth for a DL-implementation because it offers a high-quality, operationally hardened reference, and is anecdotally seen as a gold standard among forecasters.   Archives of both Level-II and Level-III products can be downloaded for free from various cloud providers, e.g., Google Cloud storage\footnote{https://cloud.google.com/storage/docs/public-datasets/nexrad.}.



\subsubsection{Peprocessing and Labelling}

 Training data requires that pairs of input velocity ($\Vf$) and target dealiased velocity ($V$) be co-aligned gate-by-gate.  
 For this reason, in an approach similar to \citet{louf++20}, training targets were created by artificially aliasing Level-III data using an input Nyquist Velocity $V_{n}$ (assumed in this work to be constant over each two-dimensional radar tilt).  
 
 The processing pipeline for creating a training pair is described as follows: Given a WSR-88D site ID and a timestamp, base Level-III velocity data (code \texttt{N0U} within the ORPG used by the Radar Operations Center) is downloaded and extracted into a two-dimensional array with a height dimension of $N_{az}=360$ azimuths and width dimension truncated to the first $N_{rng}=1152$ radial gates (this choice of radial dimension made training more convenient but is not required during testing).  The azimuthal dimension of each array is sorted such that the radial with minimum azimuth is located at the top of the array (index 0), ensuring a consistent azimuthal indexing across the training data.  Since the Level-III velocity algorithm clips data to the range of [-63.5 m s$^{-1}$, +63.5 m s$^{-1}$], regions in which the end points of this clipping range are observed need to be filled in using data from the nearest Level-II velocity field which is downsampled by averaging radials in each $1^\circ$ arc.  Note that the dealiasing of Level-II data for filling data gaps is possible since it is known that the magnitude of velocity within clipped regions exceeds the limits of Level-III data, and hence within clipped regions the Nyquist velocity can be added/subtracted until values in the region exceed the clipped value.
 
 After extracting, re-indexing, and back-filling clipped regions, the resulting two-dimensional $V$ arrays and associated Nyquist velocities are saved to disk using Tensorflow Record format\footnote{https://www.tensorflow.org/tutorials/load\_data/tfrecord} which is a simple and efficient format for training DNNs using large datasets.  At train time, a data loader reads a batch of two-dimensional dealiased velocity grids $V$ and Nyquist velocities $V_n$ and constructs a batch of artificially folded velocities $\Vf$ along with label arrays $F$ containing the target fold numbers $n$.  In order to increase the difficultly of dealiasing, the  Nyquist velocities used operationally were lowered by an amount of 10 m s$^{-1}$ (a hyper-parameter) so that the DNN would be exposed to more folded regions.  This was also done to increase the likelihood that the DNN will generalize to radars that commonly use lower Nyquist velocities compared to the S-band WSR-88D (such as radars at higher frequencies).

\begin{figure}[h]
  \centerline{\includegraphics[width=45pc]{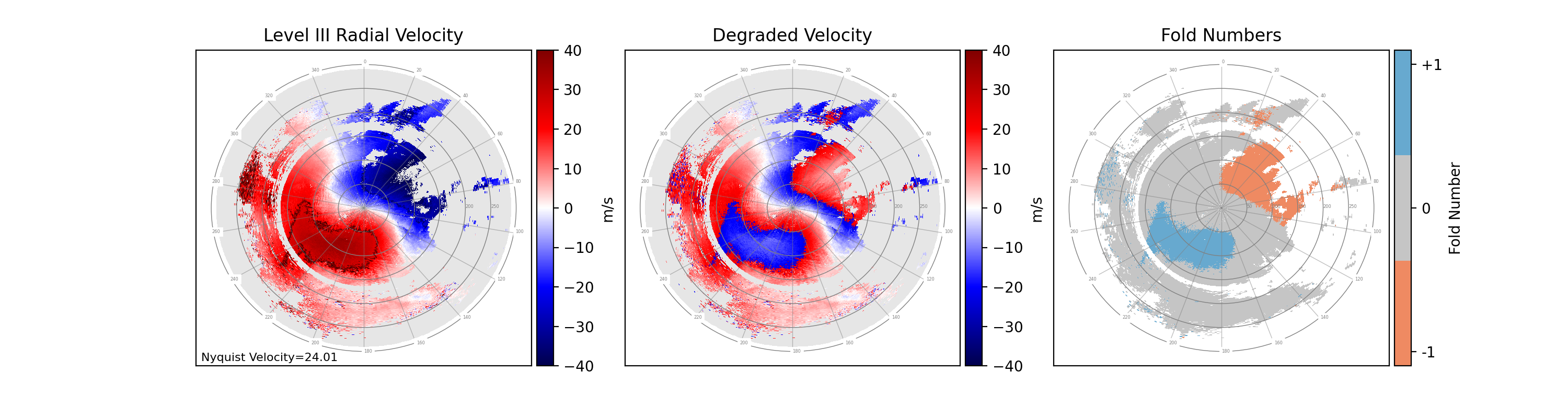}}
   \caption{Technique for creating targets for training.   Regions of Level-III velocity (left) that exceed the Nyquist velocity are identified (black contours).  These regions are aliased so that data fall in the range $[-V_{n},+V_{n}]$ (middle).   Regions that are aliased by a factor of $f$ are stored in a label mask that is used as the final training target (right).  }\label{fig:vel_degrade}
 \end{figure}

The label array $F$ used for training is of shape $N_{az} \times N_{rng} \times 6$, where the last dimension is a one-hot encoding representing the  categories shown in Table~\ref{tab:encoding}.  Note that fold values with $|n|>2$ are not included as a separate category as they are exceptionally rare in WSR-88D data.  This makes use of the algorithm in some situations andwith certain radar types less tenable, such as higher-frequency mobile radars in cases of tornadoes \citep[e.g.,][]{kurdzo++17} and hurricanes \citep[e.g.,][]{alford++19}.  However, some examples of successful unfolding with these types of radars is shown in Section 6.  Using the fold array $F$, the artificially folded array $\Vf$ was constructed by subtracting or adding the appropriate number of Nyquist velocities.  Finally, the data were padded by $12^\circ$ in the azimuthal dimension with a periodic boundary condition to create arrays with shape $N_{az}=384$.  An example of this processing can be seen in Fig. \ref{fig:vel_degrade} where the original Level-III velocity is shown on the left with regions exceeding the Nyquist velocity contoured in black.  The artificially aliased velocity is shown in the middle panel, and the resulting fold number $n$ is shown on the right.


\begin{table}[h]
    \centering
    \caption{Translation of dealiased Level-III velocity $V$ into one-hot encoded labels used for DNN training.}
    \begin{tabular}{lcc}
    \toprule
        \textbf{Range} & \textbf{Fold number} & \textbf{One-Hot Encoding}  \\
        \midrule
        Background Values & - & (1,0,0,0,0,0)    \\
         $V \le -3V_n $   & $ n=-2 $ & (0,1,0,0,0,0)\\
         $-3V_n < V \le -V_n$  & $ n=-1$ &(0,0,1,0,0,0)\\
         $-V_n < V \le V_n $  & $ n=0 $ &(0,0,0,1,0,0)\\
         $V_n < V \le 3V_n $  & $n=+1$ &(0,0,0,0,1,0)\\
         $ V \ge 3v_n$ & $n=+2  $ & (0,0,0,0,0,1)  \\
    \bottomrule
    \end{tabular}
    \label{tab:encoding}
\end{table}
 
 The result of this pipeline is four variables which are grouped as inputs: $X = (\Vf,V_n)$ and outputs: $Y=(F,V)$. These samples are grouped into mini-batches and used for DNN training.  
 
 \subsubsection{Event Selection}
 
 Selecting a training dataset that includes sufficient representation of strong radial velocities and highly sheared environments is critical for the success of the algorithm.  Storm events that contain strong and dynamic winds, such as microbursts, mesocyclones, tornadoes, and hurricanes are not only more difficult to dealias due to sudden changes in wind magnitude and direction, but also represent the greatest impact operationally.   Since these events are statistically rare, random sampling of events will likely under represent them.  In order to ensure the training and testing datasets provide adequately coverage of these events, the NCEI Storm Event Database\footnote{https://www.ncdc.noaa.gov/stormevents/} was used to identify radar sites and times where these high-impact storms were reported.  In addition to events with confirmed severe storm occurrences, a set of "null events" were also selected where either no storms were reported, or warnings were issues and no confirmed cases were surveyed.
 
 In all, the resulting dataset derived from WSR-88D was selected from 6,350 separate events from the Storm Event Database that fall between January 2017 and June 2022 (55 months).  For each of these events, several velocity images were sampled totaling 427,491 processed velocity images of shape $360 \times 1152$.  This dataset was split into three subsets: 80\% for training sampled between Jan, 2017 - Oct, 2020, 8\% for validation sampled between Nov, 2020 - May, 2021, and 12\% for testing sampled between June, 2021 - June, 2022.  Partitioning the data by time was done to avoid potential overlaps in storm events and/or radars.  When saved to disk and compressed, this combined dataset totaled 229 GB in size.
 
 \subsection{Testing Data Description}
 
In addition to the processed Level-III test data mentioned in the previous section, Level-II velocity as well as velocity data from other radars were also used for testing.  The other radars included for verification purposes are the PX-1000 X-band radar \citep{cheong++13}, and the C-band Terminal Doppler Weather Radar \citep[TDWR;][]{michelson++90,cho+10}.  PX-1000 is a transportable, solid-state, polarimetric radar that has been used to study severe weather, tornadoes, and rainfall, as well as serving as a system for radar engineering, teaching, and research.  The TDWR is an operational radar that is part of a network of 45 systems across the US, and serves as a primary tool for detecting wind shear and hazardous convection at large airports \citep{evans+89}.
 
 The characteristics of the different test sets are described in Table \ref{tab:radars}.   One challenge in using non-Level-III data is that there is no ideal dealiased reference to use for comparison, and so verification for non-Level-III derived velocity will be mostly qualitative and will utilize multiple views of the same event to increase confidence that the DNN VDA is correctly identifying the number of folds in the observed data.  Quantitative scoring will be performed only in the case of artificially aliased Level-III data, where a truth is available.
 
 \begin{table}[h]
 \caption{Summary of different radar products used for testing.}\label{tab:radars}
 \begin{center}
 \begin{tabular}{ccccc}
 \topline
  Velocity Source &  Frequency Band & Max Range (km) & Range Resolution (km) & Azimuthal Resolution ($^\circ$) \\
 \midline
  NEXRAD Level III  & S-Band & 230 & 0.25 & $1^\circ$ \\
  NEXRAD Level II  & S-Band & 230 & 0.25 & $0.5^\circ$ \\
  TDWR  & C-Band & 90 & 0.15 & $1^\circ$ \\
  PX1000  & X-Band & 60 & 0.03 & $1^\circ$ \\
 \botline
 \end{tabular}
 \end{center}
 \end{table}

\section{Model Description}\label{sec:model}

This section describes the architecture of the DNN model $f(\cdot;\theta)$.  Dealiasing is framed here as an \textit{image segmentation} task, which amounts to classifying each  pixel in a velocity image as one of $6$ categories corresponding to fold numbers $n=-2,-1,0,1,2$ or background.  After the most likely fold number is identified for each pixel, the input aliased velocity is modified by the correct scaling of Nyquist velocity using Eq. \ref{e:actv}.   

The particular DNN model chosen for this task is a U-Net \citep{ronneberger2015u}, which is a popular neural network architecture that has been applied in a number of areas including image segmentation \citep{iglovikov2018ternausnet}, biomedical image analysis \citep{haque2020deep} and image-to-image translation \citep{isola2017image}.  The U-Net has also been widely adopted in meteorological applications, including nowcasting \citep{agrawal2019machine}, precipitation estimation \citep{wang2021precipgan}, satellite image gap-filling \citep{xing2022spatiotemporal}, and several others.   What makes the U-Net an effective model is its ability to learn and process features at multiple spatial scales across an image.  This property makes the architecture ideal for dealiasing as it requires identifying sudden changes in velocities at fine scales and identifying wind patterns across the whole image without the need for environmental wind information, in order to determine the correct fold numbers. 

Another beneficial property of the U-Net architecture is that is is fully convolutional, making it possible to perform inference on images that are of a different radial shape ($N_{rng}$) compared to the training data.  Different azimuthal shapes in testing data can also be used; however, the resolution should be in 1$^\circ$ steps and it is recommended that inputs contain a full 360$^\circ$ + 12$^\circ$ of periodic boundary conditions ($N_{az}=384$).  The portability of the U-Net architecture will be demonstrated in a later section by applying the trained model to velocity data of different shapes.  It is important to note that cases with sector scans, as can be common in mobile radars, are not yet technically supported with this technique.  However, modifications can be made to the data and the algorithm in the future, including zero-padding of azimuths that are missing.  While this has not been specifically tested, it would be a straightforward approach to using the U-Net on these types of datasets.  A more-difficult problem to solve would be the limitation on 1$^\circ$ azimuthal steps, which, as of the writing of this study, will limit the algorithm's utility in some research settings.  Future work, including input from the research community, will include these types of adjustments.

\begin{figure}[h]
  \centerline{\includegraphics[width=38pc]{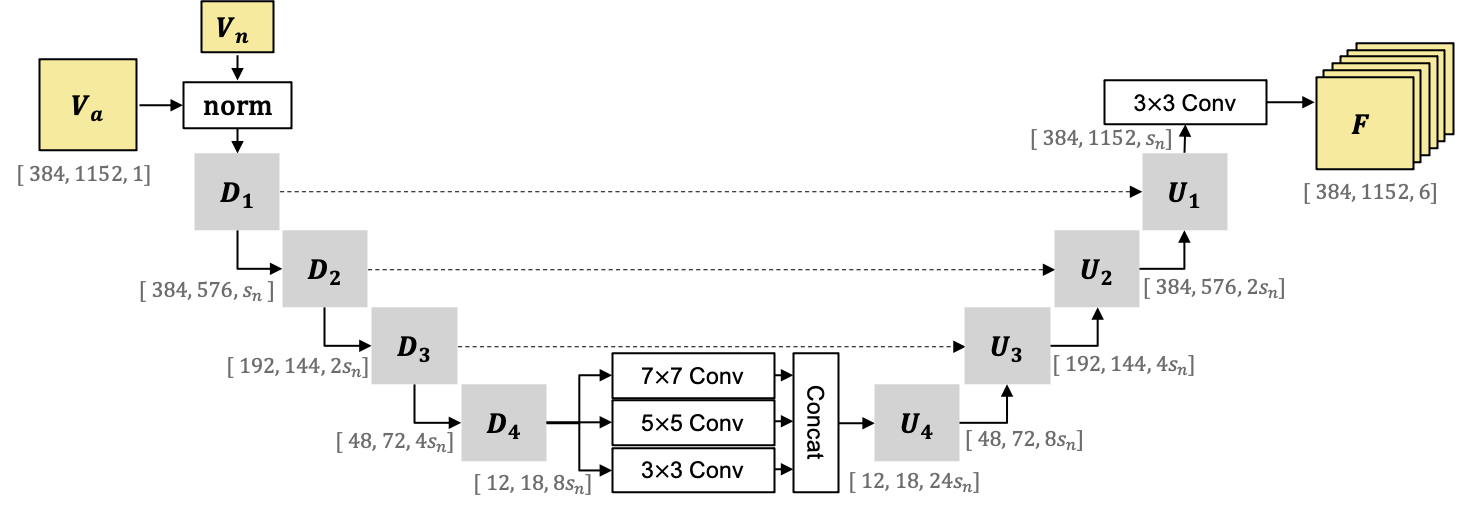}}
   \caption{ U-Net architecture for identifying aliased regions.  The inputs to the network shown on the left include a single tilt of aliased velocity $\Vf$ and the associated Nyquist velocity $V_n$.  The network is tasked with outputting the factor of $2 V_n$ required to correct the input velocity image at each gate.}\label{fig:cnn_arch}
 \end{figure}

The details of the U-Net architecture used for dealiasing are shown in Fig. \ref{fig:cnn_arch}.  The network takes two inputs.  The first input is a batch of observed (aliased) velocity $\Vf$ with shape $[B,N_{az},N_{rng},1]$, where $B$ denotes the batch size, $N_{az}$ is the (padded) azimuthal dimension, $N_{rng}$ is the radial dimension which may be of arbitrary size, and the final $1$ denotes the singleton channel dimension of the input. The second input to the model is the array of Nyquist velocities for each element of the batch used to artificially fold the target images, which has shape $[B,1]$.  The aliased velocity is initially scaled by $1/V_n$, which produces values in the interval [-1,1].   After normalization, background values in the original velocity are set to a placeholder value of -3.  

\begin{figure}[h]
  \centerline{\includegraphics[width=32pc]{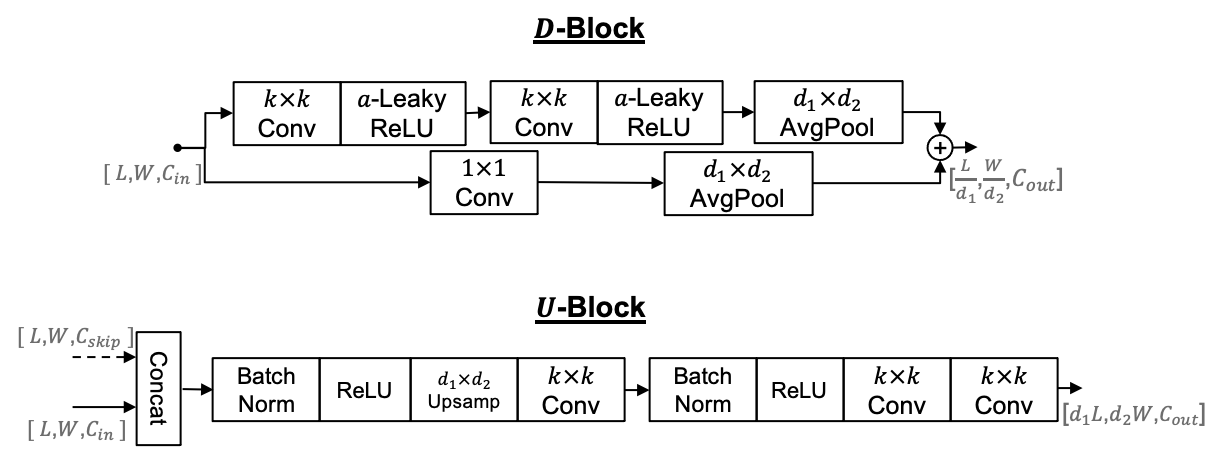}}
   \caption{ Processing blocks used in the U-Net architecture in Fig. \ref{fig:cnn_arch}.  The D-Block shown along the map transforms incoming feature maps and outputs new features at a lower resolution.  The U-Block shown along the bottom concatenates the input from the previous block and the skip connection from the parallel $D$-block.  It then applies two sets of batch normalization followed by upsampling and convolutional blocks.}\label{fig:cnn_blocks}
 \end{figure}

After normalization, the downward portion of the U-Net made up of blocks labeled $D_1,\dots,D_4$ extract features from the input, and map the data to lower and lower resolutions by cutting the sizes of each dimension by factors of 2 or 4 at each stage.  Each $D$ block is made up of a series of standard deep learning operations summarized in the top panel of Fig. \ref{fig:cnn_blocks}.  The $D$-block is comprised of two paths, the first being two two-dimensional convolutional layers with kernel size $k$ with a leaky rectified linear unit (ReLu) activation function that uses $\alpha=0.1$.  The output of the second convolution is down-sampled with a $d_{1},d_{2}$ average pooling layer.  The number of channels (features) output by each $D$ block doubles after the initial block, and the channel dimension of the first output block $D_1$ is a configurable parameter called the starting neurons or $s_n$.  The values of $k$, $d_1$, and $d_2$ for the four $D$ blocks are shown in the left portion of Table \ref{tab:hparams}.

\begin{table}[h]
\caption{Parameter settings for the U-Net blocks described in Fig. \ref{fig:cnn_blocks}}\label{tab:hparams}
\begin{center}
\begin{tabular}{cccc|cccc}
\topline
$D$-Block & $k$ & $d_1$ & $d_2$ & $U$-Block & $k$ & $d_1$ & $d_2$ \\ \hline
 $D_1$  &   7  &    1   &   2    &     $U_1$      &  3   &  4     &   4    \\ 
 $D_2$  &  5   &    2   &  4     &     $U_2$      &  3   &  4     &   2    \\ 
  $D_3$  &  3   &   4    &   2    &     $U_3$      &  3   &  2     &   4    \\ 
 $D_4$  &  3   &    4   &   4    &     $U_4$      &  3   &   1    & 2      \\ 
\botline
\end{tabular}
\end{center}
\end{table}

At the base of the dealiasing U-Net, low-resolution feature maps are processed by three parallel convolutional layers with kernel sizes of $k=$3, 5, and 7.  These additional convolutions were added to the standard U-Net architecture to maximize the receptive field of the convolutions which allows non-local features to be incorporated into the resulting segmentation.  Larger receptive fields should help correctly identify predominate winds without the need for a reference velocity field.  Experimentation with different network architectures showed that U-Net's with only a single 3x3 convolutional layer at the base produced validation losses that were approximately twice as large compared to architectures that use the parallel convolutions.  

The right portion of the U-Net is made up of four upsampling blocks $U_i$, $i=1,\dots,4$.  These blocks take two inputs.  The first input is the output of each parallel $D$-block and the second is the output of the preceding $U$-block.  The steps performed in the $U$-block are shown in the bottom panel of Fig. \ref{fig:cnn_blocks}.  The two inputs are concatenated along the channel dimension, and followed by two applications of batch normalization \citep{ioffe2015batch}, ReLU, up-sampling, and two-dimensional convolution.  In contrast to the $D$-blocks, the output of each $U$-block halves the number of features.  The values of $k$, $d_1$, and $d_2$ for the four $U$ blocks are shown in the right portion of Table \ref{tab:hparams}.  

The output of the final $U$-block is passed to a $3x3$ convolutional layer with linear activation to produce the one-hot encoded array $F$ with spatial dimensions equal to the input velocity.  The output array $F$ has 6 output channels which can be interpreted as log-odds of the six potential categories of each pixel.  The determination of the correct fold number is computed using the argmax along the channel dimension of this array.  It is worth pointing out that this output can also be combined with a softmax activation to create probabilistic predictions across different fold numbers, however this idea was not explored further.

\section{Results}\label{sec:results}

\subsection{Model Training}\label{ss:training}

The model described in Section \ref{sec:model} was implemented using the Tensorflow Keras DL framework \citep{bisong2019tensorflow} and trained on two NVIDIA Volta V100 GPUs located within the MIT Lincoln Laboratory TX-Green Super Computing Center \citep{kepner2017llgrid}.   Multiple hyper-parameter settings of the model were tested during development, and results are provided for a U-Net with starting neuron choices $s_n=8$, 16, and 24.  A categorical cross-entropy loss function was used with an Adam \citep{kingma2014adam} optimizer for weight updates.  Computations utilized mixed-precision training \citep{samsi2020compute} where network operations use 16-bit floating point, and weight updates are performed using 32-bit floating point. An initial learning rate of 0.0025 was used and was decreased by a factor of 0.99 every epoch.  Training iterations used a data distributed model with a batch size of 32 input images spread across the two GPUs.  The model was allowed to train for 200 epochs with 600 batches used per epoch. The total time for training averaged approximately 25 hours per model.

\begin{figure}[h]
  \centerline{\includegraphics[width=38pc]{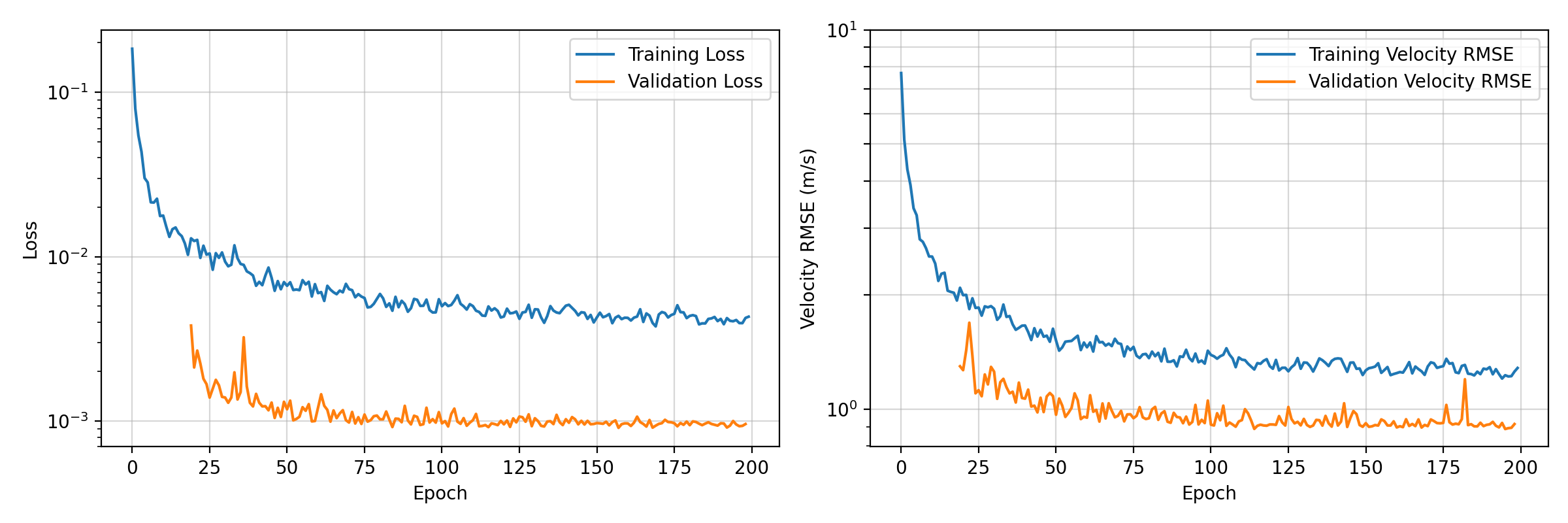}}
   \caption{ Values of the categorical cross entropy loss (left),  and RMSE of resulting dealiased velocity (right) for the training and validation set during training.  The training set uses Nyquist velocities set 10 m s${^{-1}}$ lower than the operational values for each sample, while the validation set used the actual Nyquist Velocity, which explains why the validation set shows better performance.  }\label{fig:loss_curves}
 \end{figure}

Fig. \ref{fig:loss_curves} shows the evolution of two metrics during training for a model trained with $s_n=16$.  The left panel shows categorical cross entropy loss, which describes how well the model classifies fold number across each image, and the right panel shows velocity RMSE computed between the true velocity and dealiased velocity.  The training curve shows performance on the training set which includes an artificially lowered Nyquist velocity.  To ensure accuracy on the target dataset, the validation curve reflects Nyquist velocities that were used operationally for the WSR-88D, which explains why the validation results are slightly better than training curves. As can be seen by the validation loss curves, the model seems to fully converge by around epoch 125.  The training epoch with the lowest velocity RMSE on the validation set was selected for further testing.

 \subsection{Test Set Results}\label{ss:results}

\begin{figure}[h]
  \centerline{\includegraphics[width=38pc]{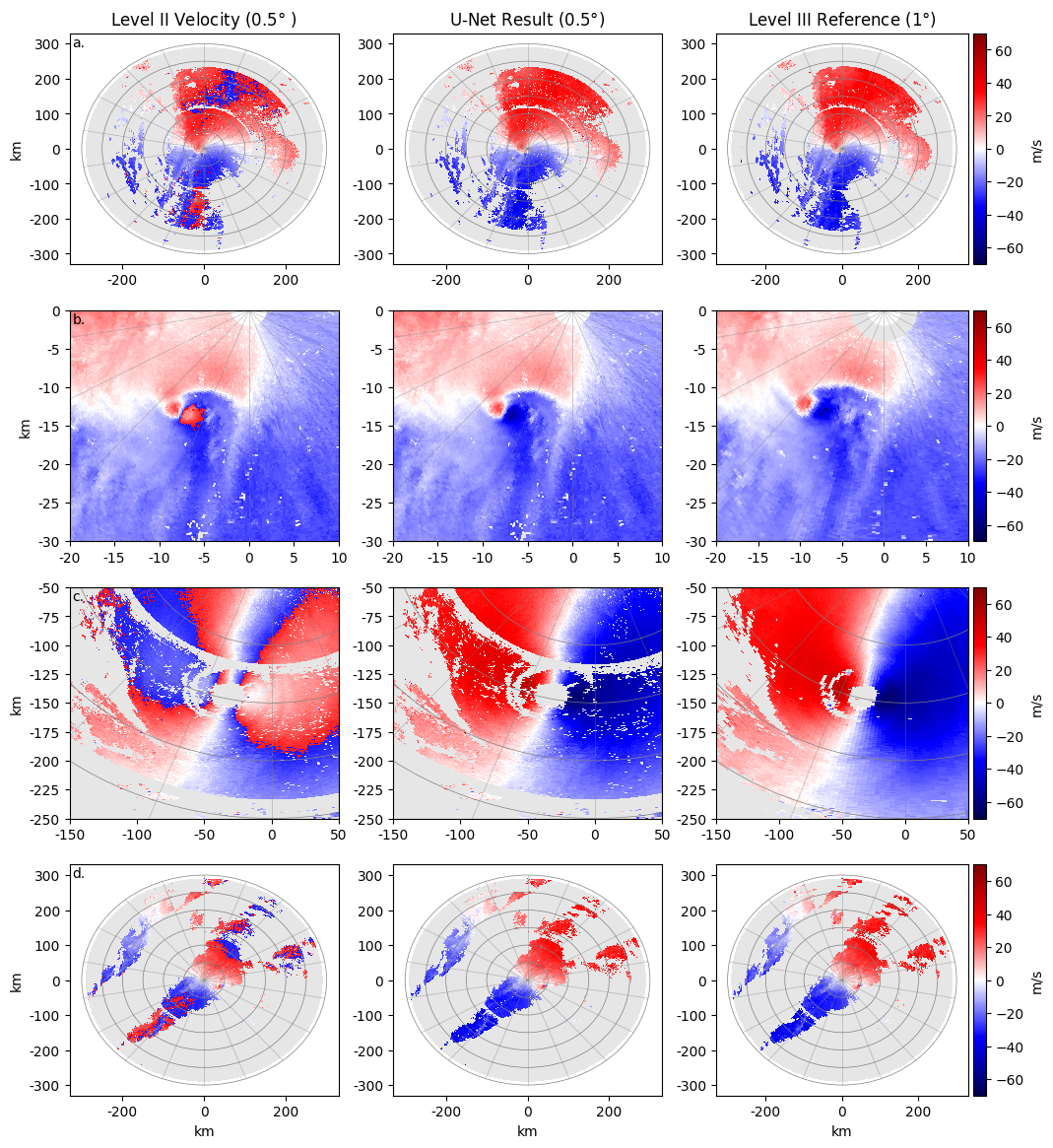}}
   \caption{ Samples of aliased Level-II velocity (left) and resulting velocity produced by the U-Net VDA.  The closest available Level-III velocity scan is also shown in the right column as a reference.}\label{fig:cases}
 \end{figure}

The trained U-Net with $s_n=16$ was tested on several instances of the $0.5^\circ$ elevation Level-II velocity data with four sample cases visualized in Fig. \ref{fig:cases}.  Note that these tilts of the Level-II velocity have a 0.5$^\circ$ azimuthal resolution which is half that of the training data.  Since the U-Net was trained on $1^\circ$ data, a single scan of Level-II V was split into a batch of two separate $1^\circ$ scans by splitting every other radial.  After both of these splits were dealiased using the U-Net, the result was recombined into a $0.5^\circ$ dataset.    

These results demonstrate the ability of the U-Net to handle challenging dealiasing situations that involve both global large-scale features, as well as resolving finer-scale aliasing in situations of high shear (e.g., severe storms).  The four cases shown include zoomed-out (case a) and zoomed-in (case b) views of a tornado near KGWX on 14 April 2019 at 0412 UTC, Hurricane Ida as it moved past KLIX on 29 August 2021 at 1639 UTC (case c), and a widespread area of convection with several isolated cells near KPAH on 11 December 2021 at 0352 UTC (case d).  The highly sheared tornadic signature in case (b), as well as the eyewall during Hurricane Ida in case (c), show examples of how the U-Net can correctly dealias areas with rapidly shifting, strong winds in complex environments.  Cases (a) and (d) show examples of how the U-Net can identify prevailing winds across an entire PPI in order to dealias isolated cells where the correct direction is not necessarily obvious without external environmental data.  In each case shown, the U-Net results are consistent with the Level-III reference image shown in the right-most column.   


For a more quantitative assessment over a larger sample size, the U-Net VDA was scored using an artificially aliased Level-III testing dataset.  In addition to loss value and average RMSE between corrected and target velocities, additional metrics of probability of detection (POD), false alarm rate (FAR), and critical success index (CSI) were also computed.  To define these metrics, a hit ($\texttt{H}$) was defined as gates where the predicted fold number $n_{pred}$ was non-zero and matched the actual fold number $n_{true}$.  Misses ($\texttt{M}$) were counted for gates where $n_{true} \neq 0$ and $n_{pred} \neq n_{true}$, and false alarms ($\texttt{F}$) were counted for gates where $n_{pred}\neq 0$ and $n_{true}=0$.  Using these counts, POD, FAR, and CSI were computed as:
\begin{equation}
    POD = \frac{\texttt{\#H}}{\texttt{\#H}+\texttt{\#M}} \quad 
    FAR = \frac{\texttt{\#F}}{\texttt{\#H}+\texttt{\#F}} \quad
    CSI = \frac{\texttt{\#H}}{\texttt{\#H}+\texttt{\#M}+\texttt{\#F}}.
\end{equation}

To serve as an additional baseline, the region-based VDA that is implemented within $\texttt{Py-ART}$ was also applied to the artificially aliased Level-III test data and scored.  According to the $\texttt{Py-ART}$ documentation\footnote{https://arm-doe.github.io/pyart/API/generated/pyart.correct.dealias\_region\_based.html}, the region-based VDA works by "finding regions of similar velocities and unfolding and merging pairs of regions until all regions are unfolded. Unfolding and merging regions is accomplished by modeling the problem as a dynamic network reduction."  Similar to the U-Net, the region-based VDA is also a two-dimensional algorithm that only requires a single tilt of velocity, which is why it was selected for comparison.  Also, environmental wind information was not supplied to the region-based algorithm, since the U-Net model does not leverage this additional data.  The default settings of the region-based algorithm were used for this study.
 
 
 
 The comparison of performance of the U-Net and region-based VDAs on the Level-III test set is shown in Table \ref{tab:test_results}.  Across all metrics, the U-Net performance exceeds that of the baseline considered, with the best performance being observed for the U-Net with the most learnable parameters ($s_n=24$), however the differences in performance metrics between the three U-Nets considered is small.  Compared to the region-based baseline, the U-Net improves the velocity RMSE by over 300\% and provides approximately 30\% increase in CSI.  A further benefit of the U-Net is its computational efficiency, running 1,200\% faster than the region-based VDA on equivalent CPU, and an impressive 9,800\% faster on GPU (which is not an option for the region-based VDA).  Also of note is the tradeoff between skill and computation time for different $s_n$.  Larger models also require additional memory to run, which may limit the ability to use these models on certain hardware.
 
   \begin{table}[h]
 \caption{Performance of the U-Net VDA for different network sizes (controlled by the hyperparamter $s_n$) on the Level-III test set.  Results from the region-based VDA implemented in the $\texttt{Py-ART}$ library are also shown as a baseline.  The U-Net VDA scores high across all metrics, and shows small increases as the number of parameters in the model increases.  The run-times of different VDAs are shown, illustrating the computational speed of the DL approach compared to the baseline. }\label{tab:test_results}
 \begin{center}
 \begin{tabular}{ccccccc}
 \topline
  VDA &  RMSE$\downarrow$ & POD$\uparrow$ & FAR$\downarrow$ & CSI$\uparrow$ & CPU Runtime (ms/scan)$\downarrow$ & GPU Runtime (ms/scan)$\downarrow$\\
 \midline
 U-Net ($s_n=8$) & 1.6387 & 0.9827 & \textbf{0.0105} & 0.9726 & \textbf{70.6} & \textbf{16.4} \\
  U-Net ($s_n=16$) & 1.5300 & 0.9878 & 0.0117 & 0.9764 & 129.3 & 17.4 \\
  U-Net ($s_n=24$) & \textbf{1.4753} & \textbf{0.9887} & 0.0109 & \textbf{0.9781} & 206.3 & 24.0 \\
  Region-based (\texttt{Py-ART})  & 5.6351 & 0.9045 & 0.1887 & 0.7472 & 1669 & -  \\
 \botline
 \end{tabular}
 \end{center}
 \end{table}
 
\begin{figure}[h]
  \centerline{\includegraphics[width=38pc]{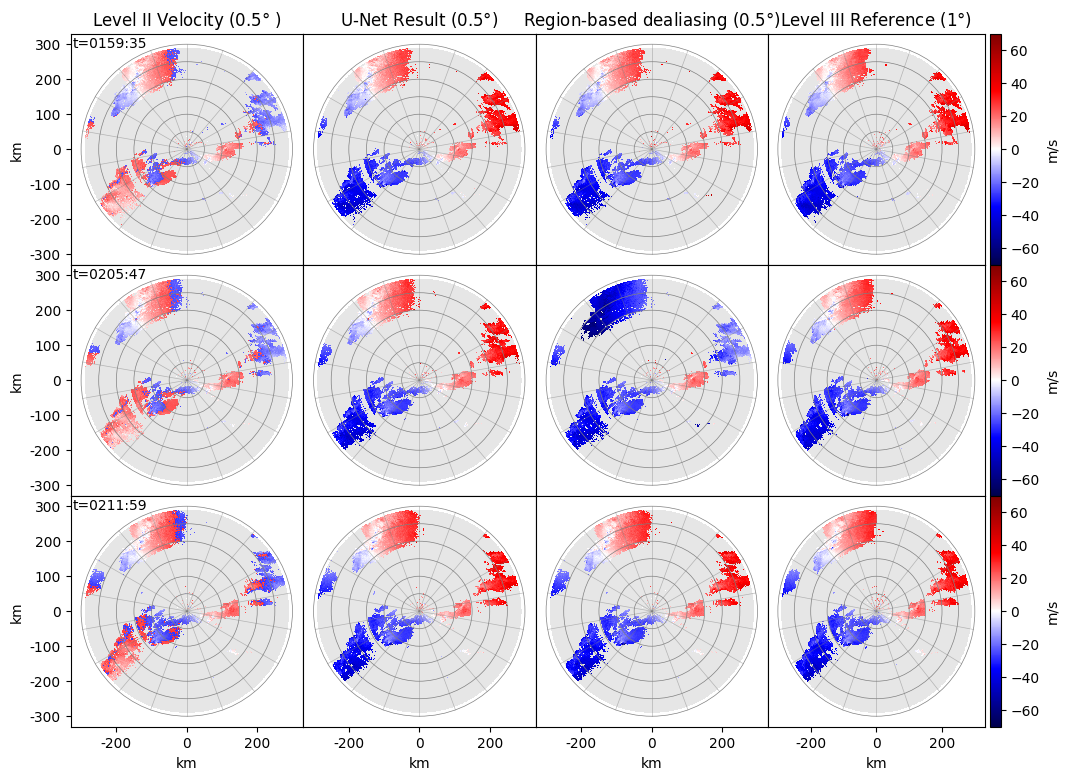}}
   \caption{ A comparison of the U-Net VDA with the region-based VDA implemented in $\texttt{Py-ART}$. The three rows represent consecutive time steps of velocity from KPAH on 11 December 2021. In the middle time step, the region-based VDA incorrectly dealiases isolated storms in the northern portions of the scan.}\label{fig:pyart_fail}
 \end{figure}

 It should be noted that comparisons of skill scores between the U-Net and region-based VDAs reported in Table \ref{tab:test_results} should be interpreted with some care, as the U-Net has the advantage of being trained to emulate the same algorithm used to create the test set.  Since the ORPG VDA utilizes environmental wind data, it is reasonable to assume that its output will be more accurate, on average, than the region-based baseline that was not supplied this additional information.  However, it is possible that the U-Net learned and reproduced the same mistakes as the ORPG, leading to overly optimistic skill scores in this case.  In order to explore this possibility further, several case studies where the ORPG and region-based VDAs disagreed were examined by human experts to judge which are correct.  In virtually all cases that were examined, results similar to the example shown in Fig. \ref{fig:pyart_fail} were observed.  Three consecutive frames of Level-II velocity (left column) are dealiased by the U-Net and region-based algorithms (second and third columns).  In the second row, without providing environmental wind information, the region-based algorithm produces incorrect results for isolated regions in the north and northeast portions of the scan, which disagrees with both the U-Net and Level-III reference (right column), which provide consistent results across all frames.

 \subsubsection{Case Studies}\label{ss:other_radars}
 
 This section considers two case studies comprised of tornadic storms that were simultaneously sampled by multiple radar systems. As such, each radar provides a unique perspective of the storm velocity structures, affected by its unique system's aliasing characteristics. Application of U-Net VDA with $s_n=16$ in these cases provides intercomparison between the resulting dealiased wind fields, and highlights the portability of this algorithm across S-, C-, and X-band weather radar systems.  Note that no additional training of the U-Net was performed prior to applying the algorithm to these other radars.
 
\begin{figure}[h]
  \centerline{\includegraphics[width=38pc]{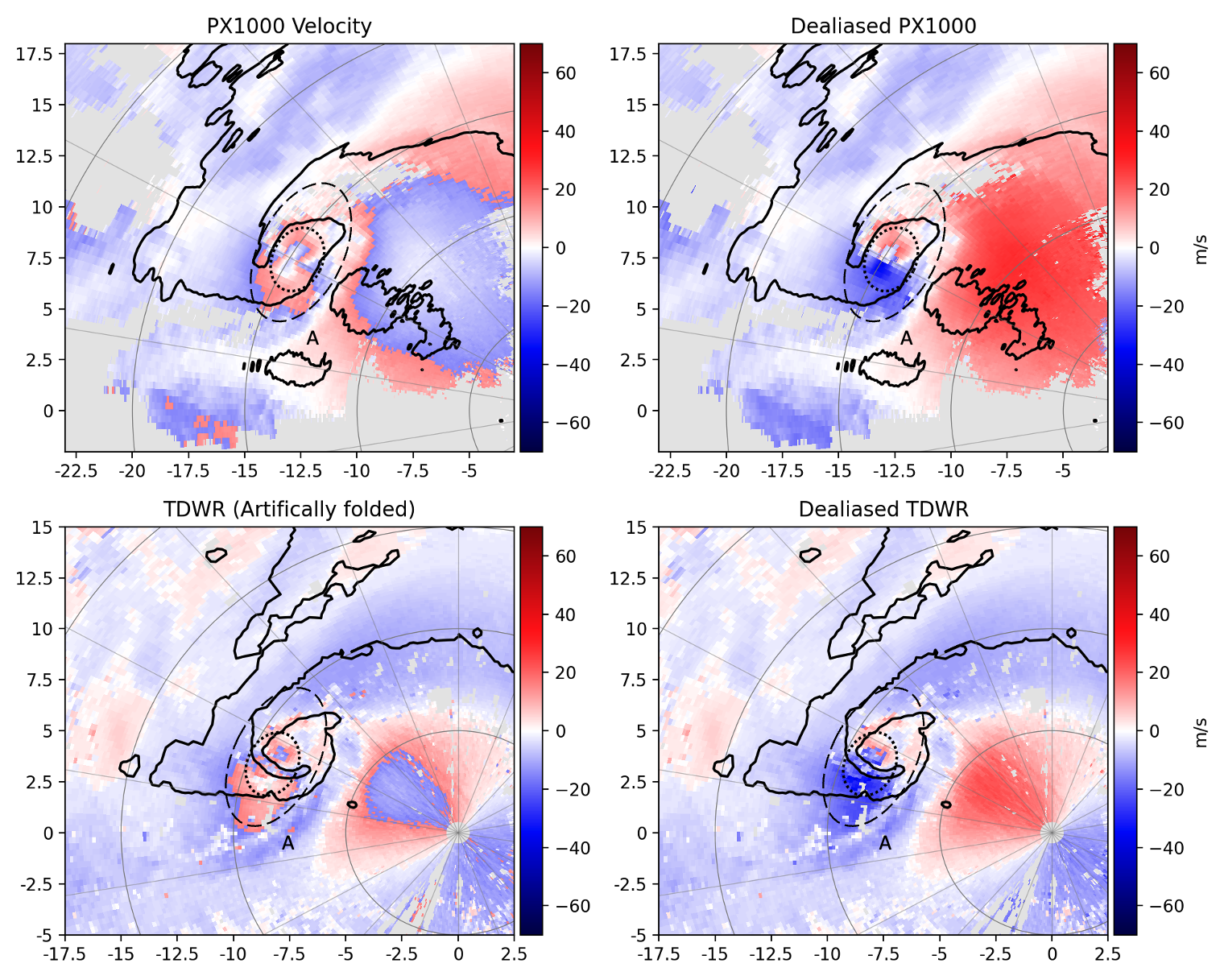}}
   \caption{Application of the U-Net algorithm to the PX-1000 radar for the 20 May 2013 EF5 tornado in Moore, OK at 2001:44 UTC (PX-1000) and 2002:00 UTC (TDWR).  The solid line denotes a smoothed 30-dBZ reflectivity factor contour, the dotted line indicates the location of the tornado, the dashed line indicates the approximate mesocyclone location, and the ``A'' label denotes the leading edge of the rear flank gust front.  The broad-scale mesocyclone is dealiased with relatively high accuracy, but the core tornadic circulation contains errors.  The axis labels are equivalent to the distance from each respective radar in km.}\label{fig:moore}
 \end{figure}

 The first case is an EF5 tornado that impacted Moore, Oklahoma on 20 May 2013, as observed by the University of Oklahoma PX-1000 research radar and the Oklahoma City TDWR \citep{kurdzo++15a}.  The selected PX-1000 time is 2001:44 UTC, while the TDWR time is 2002:00 UTC.  At these times, the differential velocity within the tornado via PX-1000 was approximately 65 m s$^{-1}$ \citep{kurdzo++15a} and the tornado was producing EF4 damage in western Newcastle \citep{burgess++14,atkins++14}.  When compared to the manual dealiasing in Fig. 2a-b in \citet{kurdzo++15a}, the U-Net VDA accurately dealiases the broad-scale mesocyclone (Fig. \ref{fig:moore}).  The VDA on TDWR data also performs qualitatively well in the mesocyclone near the rear flank gust front where there are strong inbound radial velocities.  The VDA on both radars also accurately dealiases the inflow to the supercell (areas of outbounds/red colors northwest of each radar).
 
 Within the tornadic core circulation, the VDA has difficulty dealiasing velocities properly.  There are several reasons that can explain these difficulties.  First, it should be noted that dealiasing in high-shear environments such as tornadoes is exceptionally difficult, as recently described by \citet{feldmann++20}.  This is particularly true at high frequencies due to the lower Nyquist velocity.  Additionally, when the Nyquist velocity is low, the wide spectrum width in tornadoes makes it difficult for the spectral estimator to determine an accurate velocity \citep{yu++07a}.  In Fig. \ref{fig:moore}, a qualitative assessment shows that the VDA applied to PX-1000 is not able to dealias velocities within the core circulation, specifically in the region of outbounds (red colors).  This is likely due to the highest area of spectrum width being in that region, but may also be due to the existence of three folds, which this iteration of the VDA is not capable of unfolding.  The VDA applied to TDWR performs almost identically, with the area in the rear flank (inbounds/blue colors) dealiased properly, but the core circulation dealiased improperly.  In the case of the TDWR, the core circulation is improperly dealiased as inbounds rather than outbounds.
 
 It is important to note that for the TDWR, the velocity estimates were properly dealiased in the raw data due to the dual-PRT VDA on the operational TDWR radar signal processor \citep{cho10}.  This dual-PRT approach had an aliasing velocity over 40 m s$^{-1}$.  In order to produce the images in Fig. \ref{fig:moore}, an artificial Nyquist velocity of 16.5 m/s was applied to the raw velocity data to simulate aliased velocities.


\begin{figure}[h]
  \centerline{\includegraphics[width=38pc]{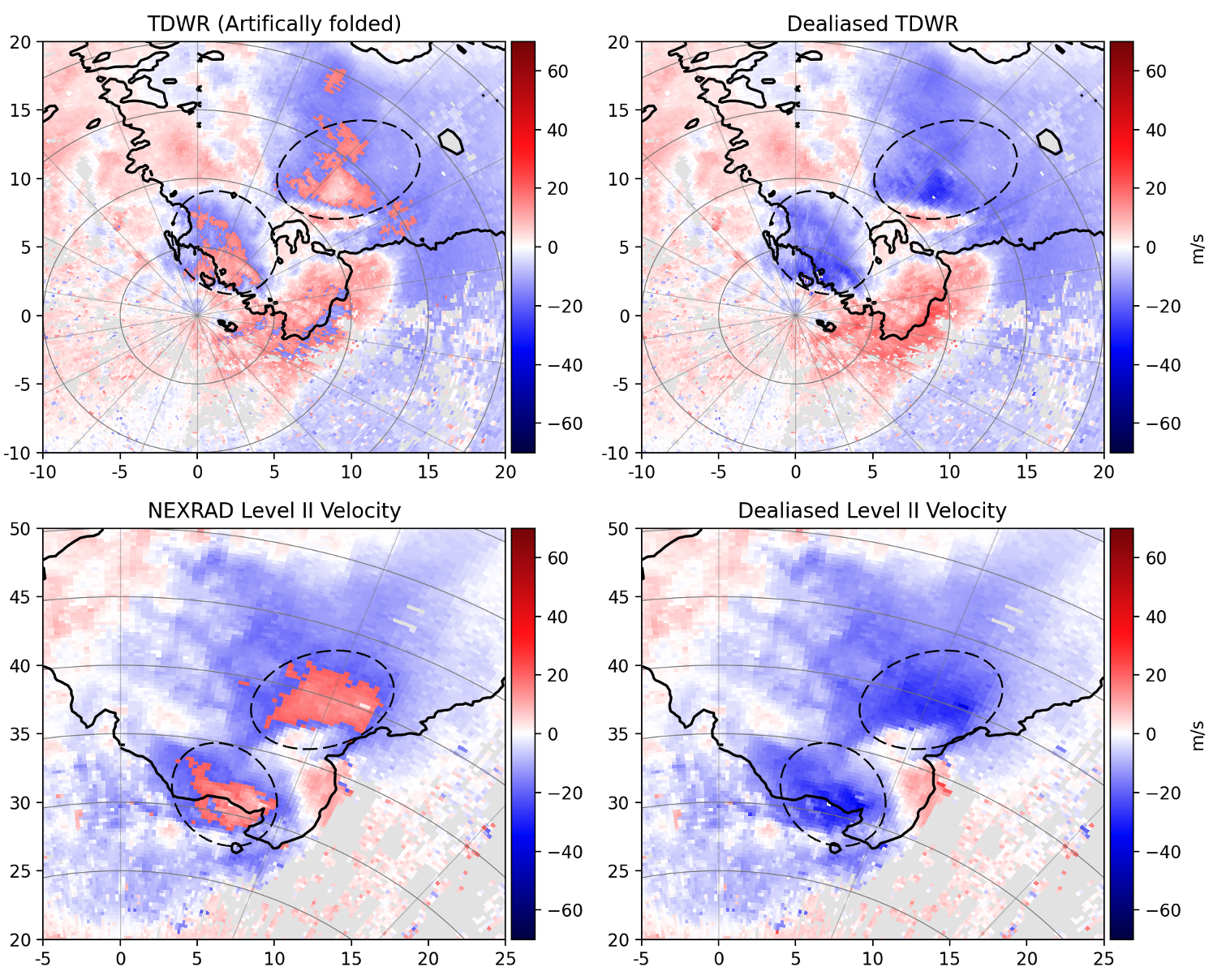}}
   \caption{Application of the U-Net algorithm to two different test datasets: Artificially folded TDWR data on 14 June 2022 at 0003 UTC (top row) and NEXRAD Level-II velocity on the same date at 0001 UTC (bottom row).  The solid line denotes a smoothed 30-dBZ reflectivity factor contour, and the dashed lines indicate the primary areas of velocity aliasing that were recovered.  The U-Net VDA adequately corrects the aliased velocities in both cases, with a small area of errors near (just north of) the TDWR.  The axis labels are equivalent to the distance from each respective radar in km.}\label{fig:june22}
 \end{figure}

The second case is a long-lived supercell that moved across northern Illinois and Indiana on the evening of 13 June 2022, producing two brief EF0 tornadoes in the Chicago suburbs.\footnote{https://www.weather.gov/lot/2022jun13}  The post-tornadic storm is shown in Fig. \ref{fig:june22} as it passed through an area located along approximately the same radial direction from two nearby radars, a TDWR (TORD; Fig. \ref{fig:june22} top row) and a WSR-88D (KLOT; Fig. \ref{fig:june22} bottom row), producing comparable radial velocity perspectives. Across both systems, the U-Net VDA performs qualitatively well, with two minor regions of errors when dealiasing the TDWR at ranges close to the radar (Fig. \ref{fig:june22} top row). The WSR-88D is dealiased without any apparent errors (Fig. \ref{fig:june22} bottom row).  


\section{Discussion, Future Directions, and Conclusions}\label{sec:discussion}

The results in Section \ref{sec:results} demonstrate that DL has potential to be an effective algorithmic tool for weather radar product development.  A major challenge for supervised learning approaches like the one taken here is that they require large quantities of ``truth'' in order to train.  In many cases (particularly with weather radar), large quantities of truth are rarely available, so if DL techniques are to be widely adopted in this space, alternative sources and approaches need to be leveraged.  

In the case of dealiasing, this issue was overcome by utilizing an operational algorithm as truth, namely the Level-III velocity product dealiased by the ORPG VDA.  As a result, the DNN trained here can be viewed as a kind of ML-"surrogate" to an already operational algorithm which is capable of generating very similar output (CSIs > 95\% and RMSEs near 1.5 m s$^{-1}$) in a fraction of the run-time, and without access to environmental wind information.  These faster implementations enabled by the DL approach could be leveraged in real-time settings where traditional dealiasing techniques fail to keep up with certain rapid scanning strategies.  For example, if real-time dealiasing were desired for an extremely rapidly updating research radar (e.g., as fast as 2-s updates such as presented in \citet{pazmany++13}), this implementation of a VDA would be capable of keeping up in real-time with relatively modest hardware. The DL models may also be useful in research settings by allowing massive velocity datasets to be dealiased in minutes rather than hours or days using traditional approaches.  Finally, more-complex VDAs such as higher-dimensional techniques could be trained using DL in the future in order to utilize their approaches in real-time.


Another advantage of the ORPG surrogates is portability.  Modern DL frameworks and deployment frameworks simplify the importing and application of trained models to new datasets without the need for additional software installations and dependencies.  In addition, the ORPG code is written exclusively for the WSR-88D network, with little ability to port algorithms to other types of radar (i.e., different vendors, frequencies, beamwidths, etc.).   This flexibility provided by the U-Net allows for simpler application of an ORPG-like VDA to new radar types without the need for significant data reformatting required by the traditional ORPG algorithm.  Improved performance on other radar types was also gained by training with Nyquist velocities that were lower than what is commonly set by WSR-88D operators.  This enabled impressive performance on other unseen radars (TDWR, PX-1000) despite the fact that the model wasn't trained directly on these systems. The prospect of porting ORPG-like algorithms to other radar systems may provide benefits to weather radar networks outside of the USA where these algorithms are not currently available. Further performance improvements on non-WSR-88D data can potentially be enabled using transfer learning by fine-tuning the network developed here.  

It should be noted that the VDA is far from the only ORPG product that is computationally complex.  As of Build 20, the ORPG maintains a suite of over 150 products, which must all be run in real-time and keep up with the fastest scan rates.  This plethora of algorithms may need to be run in a much-more expeditious manner in the future with faster scan rates, such as those discussed in \citet{kollias++22} and \citet{palmer++22}.  Additionally, other radar systems may not have access to the hardware available for the ORPG.  For example, research radars may want to implement WSR-88D algorithms in real-time but do not have the computational capabilities.

Given the results from this work, a natural question to ask is how this approach can be leveraged for other weather radar product enhancements.   A whole suite of Level-III radar products have been successfully developed using traditional signal, image, and video processing techniques, and these methods have their constraints and weaknesses that can lead to difficulty in both implementation and operations.    Additionally, like any algorithm, many of the ORPG algorithms suffer from false alarms, missed detections, and/or relatively noisy interpretations of un-truthed data \citep[e.g., the hydrometeor classification algorithm, or HCA;][]{park++09}.  However, ORPG algorithms, in many cases, are considered state-of-the-art, and are desirable tools to have available to other radar systems.  One way to explore this possibility is to implement other ML-based surrogates to ORPG algorithms.  ML-based versions of the ORPG algorithms could result in several advantages, ranging from faster compute times, to (potentially) improved accuracy through continued fine-tuning of ML-surrogates using additional datasets. 


Overall, in this work, a DL approach was developed for two-dimensional velocity dealiasing of weather radar data.  To construct a training dataset, archives of Level-II and Level-III WSR-88D radial velocity products were combined to create a large dataset of over 100,000 dealiased velocity PPIs.  To create aliased/dealiased pair samples for training, these scans were artificially folded using a known Nyquist velocity.  

A U-Net model was trained using this dataset to output the correct fold number at each point in an input velocity image.  The U-Net model developed here utilized several convolutional layers of different sizes to ensure it properly learned both global and local structure in the input image.  Once trained, the U-Net model generated similar results compared to the operational ORPG algorithm from which is was trained based on a test set of held-out WSR-88D data. Encouragingly, the model also performed reasonably well on radars that were not included in the training dataset, including the TDWR and PX-1000 radars.

Future work will continue to explore potential DL applications in radar meteorology, including the development of other ORPG surrogate algorithms.  These techniques have the potential to improve existing radar products and to enable ORPG product generation on non-WSR-88D radars.

\acknowledgments
The authors would like to thank Mike Donovan, Betty Bennett, and Mike Istok for helpful discussions regarding the WSR-88D ORPG. The authors would also like to thank the anonymous reviewers for their valuable insights and suggestions. DISTRIBUTION STATEMENT A. Approved for public release: distribution unlimited. This material is based upon work supported by the Department of the Air Force under Air Force Contract No. FA8702-15-D-0001. Any opinions, findings, conclusions or recommendations expressed in this material are those of the author(s) and do not necessarily reflect the views of the Department of the Air Force.

\datastatement
Trained model weights, sample data, and code for running the U-Net VDA can be found on https://github.com/mit-ll/unet-vda.  Source code for ORPG can be downloaded from https://www.weather.gov/code88d/.  WSR-88D data used in this work is publicly available from AWS Open Data Registry\footnote{https://registry.opendata.aws/noaa-nexrad/} and Google Cloud Storage\footnote{https://cloud.google.com/storage/docs/public-datasets/nexrad}.  TDWR data is publicly available from the National Center for Environmental Information\footnote{https://www.ncei.noaa.gov/products/radar/terminal-doppler-weather-radar}.  The PX-1000\footnote{https://arrc.ou.edu/radar\_px1000.html} is maintained and operated by the Advanced Radar Research Center (ARRC) of the University of Oklahoma and data can be requested by contacting data@arrc.ou.edu.

\bibliographystyle{ametsocV6}

\bibliography{Kurdzo_Bibliography.bib,references.bib}

\begin{thebibliography}{72}
\providecommand{\natexlab}[1]{#1}
\providecommand{\url}[1]{\texttt{#1}}
\renewcommand{\UrlFont}{\rmfamily}
\providecommand{\urlprefix}{URL }
\expandafter\ifx\csname urlstyle\endcsname\relax
  \providecommand{\doi}[1]{https://doi.org/\discretionary{}{}{}#1}\else
  \providecommand{\doi}{https://doi.org/\discretionary{}{}{}\begingroup
  \urlstyle{rm}\Url}\fi
\providecommand{\eprint}[2][]{\url{#2}}

\bibitem[{Agrawal et~al.(2019)Agrawal, Barrington, Bromberg, Burge, Gazen,, and
  Hickey}]{agrawal2019machine}
Agrawal, S., L.~Barrington, C.~Bromberg, J.~Burge, C.~Gazen, and J.~Hickey,
  2019: Machine learning for precipitation nowcasting from radar images.
  \textit{arXiv preprint arXiv:1912.12132}.

\bibitem[{Alford et~al.(2019)Alford, Biggerstaff,, and Carrie}]{alford++19}
Alford, A.~A., M.~I. Biggerstaff, and G.~D. Carrie, 2019: {Mobile ground-based
  SMART radar observations and wind retrievals during the landfall of Hurricane
  Harvey (2017)}. \textit{Geosci. Data J.}, \textbf{6~(2)}, 205--213,
  \doi{10.1002/gdj3.82}.

\bibitem[{Atkins et~al.(2014)Atkins, Butler, Flynn,, and Wakimoto}]{atkins++14}
Atkins, N.~T., K.~M. Butler, K.~R. Flynn, and R.~M. Wakimoto, 2014: {An
  integrated damage, visual, and radar analysis of the 2013 Moore Oklahoma EF5
  tornado}. \textit{Bull. Amer. Meteor. Soc.}, \textbf{95~(10)}, 1549--1561.

\bibitem[{Biggerstaff et~al.(2021)Biggerstaff, Alford, Carrie,, and
  Stevenson}]{biggerstaff++21}
Biggerstaff, M.~I., A.~A. Alford, G.~D. Carrie, and J.~A. Stevenson, 2021:
  {Hurricane Florence (2018): Long duration single- and dual-Doppler
  observations and wind retrievals during landfall}. \textit{Geosci. Data J.},
  \doi{10.1002/gdj3.137}.

\bibitem[{Bisong(2019)}]{bisong2019tensorflow}
Bisong, E., 2019: Tensorflow 2.0 and keras. \textit{Building Machine Learning
  and Deep Learning Models on Google Cloud Platform}, Springer, 347--399.

\bibitem[{Bluestein et~al.(2007)Bluestein, French, Tanamachi, Frasier,
  Hardwick, Junyent,, and Pazmany}]{bluestein++07}
Bluestein, H.~B., M.~M. French, R.~L. Tanamachi, S.~Frasier, K.~Hardwick,
  F.~Junyent, and A.~L. Pazmany, 2007: {Close-range observations of tornadoes
  in supercells made with a dual-polarization, X-band, mobile Doppler radar}.
  \textit{Mon. Wea. Rev.}, \textbf{135~(4)}, 1522--1543.

\bibitem[{Brown et~al.(1978)Brown, Lemon,, and Burgess}]{brown++78}
Brown, R.~A., L.~R. Lemon, and D.~W. Burgess, 1978: {Tornado detection by
  pulsed Doppler radar}. \textit{Mon. Wea. Rev.}, \textbf{106}, 29--38.

\bibitem[{Burgess et~al.(2014)}]{burgess++14}
Burgess, D., and Coauthors, 2014: {20 May 2013 Moore, Oklahoma tornado: Damage
  survey and analysis}. \textit{Wea. Forecasting}, \textbf{29~(5)}, 1229--1237.

\bibitem[{Burgess and Lemon(1990)Burgess, and Lemon}]{burgess+90}
Burgess, D.~W., and L.~R. Lemon, 1990: {Severe thunderstorm detection by
  Radar}. \textit{Radar in Meteorology}, Springer, 619--647.

\bibitem[{Burgess et~al.(2002)Burgess, Magsig, Wurman, Dowell,, and
  Richardson}]{burgess++02}
Burgess, D.~W., M.~A. Magsig, J.~Wurman, D.~C. Dowell, and Y.~Richardson, 2002:
  {Radar observations of the 3 May 1999 Oklahoma City tornado}. \textit{Wea.
  Forecasting}, \textbf{17~(3)}, 456--471.

\bibitem[{Chandrasekar(2020)}]{chandra20}
Chandrasekar, V., 2020: {AI in weather radars}. 2020 IEEE Radar Conference
  (RadarConf20), Vol.~00, 1--3, \doi{10.1109/radarconf2043947.2020.9266442}.

\bibitem[{Chen et~al.(2020)Chen, Chandrasekar, Cifelli,, and Xie}]{chen++19}
Chen, H., V.~Chandrasekar, R.~Cifelli, and P.~Xie, 2020: {A machine learning
  system for precipitation estimation using satellite and ground radar network
  observations}. \textit{IEEE Trans. on Geoscience and Remote Sensing},
  \textbf{58~(2)}, 982--994, \doi{10.1109/tgrs.2019.2942280}.

\bibitem[{Cheong et~al.(2013)Cheong, Kelley, Palmer, Zhang, Yeary,, and
  Yu}]{cheong++13}
Cheong, B.~L., R.~Kelley, R.~D. Palmer, Y.~Zhang, M.~Yeary, and T.-Y. Yu, 2013:
  {PX-1000: A solid-state polarimetric X-band weather radar and time-frequency
  multiplexed waveform for blind range mitigation}. \textit{IEEE Trans.
  Instrum. Meas.}, \textbf{62~(11)}, 3064--3072.

\bibitem[{Cho(2010)}]{cho10}
Cho, J. Y.~N., 2010: {Signal processing algorithms for the Terminal Doppler
  Weather Radar: Build 2}. Tech. rep., Project Rep. ATC-363, MIT Lincoln
  Laboratory, Lexington, MA, 92 pp.

\bibitem[{Cho and Weber(2010)Cho, and Weber}]{cho+10}
Cho, J. Y.~N., and M.~E. Weber, 2010: {Terminal Doppler Weather Radar
  enhancements}. \textit{2010 IEEE Radar Conference}, 1245--1249,
  \doi{10.1109/radar.2010.5494427}.

\bibitem[{CODE(2021)}]{code21}
CODE, 2021: {WSR-88D CODE}. \urlprefix\url{https://www.weather.gov/code88d/}.

\bibitem[{Collis et~al.(2018)}]{collis++18}
Collis, S., and Coauthors, 2018: {It's time for color vision deficiency
  friendly color maps in the radar community}. \textit{European Radar Conf.},
  \urlprefix\url{https://ntrs.nasa.gov/api/citations/20180004634/downloads/20180004634.pdf}.

\bibitem[{Doviak and Zrnic(1993)Doviak, and Zrnic}]{doviak+93}
Doviak, R.~J., and D.~S. Zrnic, 1993: \textit{{Doppler Radar and Weather
  Observations}}. Dover Publications.

\bibitem[{Eilts and Smith(1990)Eilts, and Smith}]{eilts+90}
Eilts, M., and S.~Smith, 1990: {Efficient dealiasing of Doppler velocities
  using local environment constraints}. \textit{J. Atmos. Oceanic Technol.},
  \textbf{7}, 118--128.

\bibitem[{Evans and Turnbull(1989)Evans, and Turnbull}]{evans+89}
Evans, J., and D.~Turnbull, 1989: {Development of an automated windshear
  detection system using Doppler weather radar}. \textit{Proceedings of the
  IEEE}, \textbf{77}, 1661--1673.

\bibitem[{Feldmann et~al.(2020)Feldmann, James, Boscacci, Leuenberger, Gabella,
  Germann, Wolfensberger,, and Berne}]{feldmann++20}
Feldmann, M., C.~N. James, M.~Boscacci, D.~Leuenberger, M.~Gabella, U.~Germann,
  D.~Wolfensberger, and A.~Berne, 2020: {R2D2: A region-based recursive Doppler
  dealiasing algorithm for operational weather radar}. \textit{J. Atmos.
  Oceanic Technol.}, \textbf{37~(12)}, 2341--2356,
  \doi{10.1175/jtech-d-20-0054.1}.

\bibitem[{Haque and Neubert(2020)Haque, and Neubert}]{haque2020deep}
Haque, I. R.~I., and J.~Neubert, 2020: Deep learning approaches to biomedical
  image segmentation. \textit{Informatics in Medicine Unlocked}, \textbf{18},
  100\,297.

\bibitem[{He et~al.(2012)He, Li, Zou,, and Ray}]{he++12}
He, G., G.~Li, X.~Zou, and P.~S. Ray, 2012: {A velocity dealiasing scheme for
  synthetic C-band data from China's new generation weather radar system
  (CINRAD)}. \textit{J. Atmos. Oceanic Technol.}, \textbf{29~(9)}, 1263--1274,
  \doi{10.1175/jtech-d-12-00010.1}.

\bibitem[{Helmus and Collis(2016)Helmus, and Collis}]{helmus+16}
Helmus, J.~J., and S.~M. Collis, 2016: {The Python ARM radar toolkit (Py-ART),
  a library for working with weather radar data in the Python programming
  language}. \textit{J. Open Res. Softw.}, \textbf{4~(1)}, e25,
  \doi{10.5334/jors.119}.

\bibitem[{Hilburn et~al.(2021)Hilburn, Ebert-Uphoff,, and
  Miller}]{hilburn2021development}
Hilburn, K.~A., I.~Ebert-Uphoff, and S.~D. Miller, 2021: Development and
  interpretation of a neural-network-based synthetic radar reflectivity
  estimator using {GOES-R} satellite observations. \textit{J. Appl. Meteor.
  Climatol.}, \textbf{60~(1)}, 3--21.

\bibitem[{Hong and Gourley(2014)Hong, and Gourley}]{hong+15}
Hong, Y., and J.~J. Gourley, 2014: \textit{{Radar Hydrology: Principles,
  Models, and Applications}}. CRC Press.

\bibitem[{Husnoo et~al.(2021)Husnoo, Darlington, Torres,, and
  Warde}]{husnoo++21}
Husnoo, N., T.~Darlington, S.~Torres, and D.~Warde, 2021: {A neural-network
  quality control scheme for improved quantitative precipitation estimation
  accuracy on the UK weather radar network}. \textit{J. Atmos. Oceanic Tech.},
  \doi{10.1175/jtech-d-20-0120.1}.

\bibitem[{Iglovikov and Shvets(2018)Iglovikov, and
  Shvets}]{iglovikov2018ternausnet}
Iglovikov, V., and A.~Shvets, 2018: Ternausnet: U-net with {VGG11} encoder
  pre-trained on imagenet for image segmentation. \textit{arXiv preprint
  arXiv:1801.05746}.

\bibitem[{Ioffe and Szegedy(2015)Ioffe, and Szegedy}]{ioffe2015batch}
Ioffe, S., and C.~Szegedy, 2015: Batch normalization: Accelerating deep network
  training by reducing internal covariate shift. \textit{International
  Conference on Machine Learning}, PMLR, 448--456.

\bibitem[{Isola et~al.(2017)Isola, Zhu, Zhou,, and Efros}]{isola2017image}
Isola, P., J.-Y. Zhu, T.~Zhou, and A.~A. Efros, 2017: Image-to-image
  translation with conditional adversarial networks. \textit{Proceedings of the
  IEEE Conference on Computer Vision and Pattern Recognition}, 1125--1134.

\bibitem[{James and Houze(2001)James, and Houze}]{james+01}
James, C.~N., and R.~A. Houze, 2001: {A real-time four-dimensional Doppler
  dealiasing scheme}. \textit{J. Atmos. Oceanic Tech.}, \textbf{18~(10)},
  1674--1683, \doi{10.1175/1520-0426(2001)018<1674:artfdd>2.0.co;2}.

\bibitem[{Jatau et~al.(2021)Jatau, Melnikov,, and Yu}]{jatau++21}
Jatau, P., V.~Melnikov, and T.-Y. Yu, 2021: {A machine learning approach for
  classifying bird and insect radar echoes with S-band polarimetric weather
  radar}. \textit{J. Atmos. Oceanic Tech.}, \doi{10.1175/jtech-d-20-0180.1}.

\bibitem[{Jing and Wiener(1993)Jing, and Wiener}]{jing+93}
Jing, Z., and G.~Wiener, 1993: {Two-dimensional dealiasing of Doppler
  velocities}. \textit{J. Atmos. Oceanic Tech.}, \textbf{10~(6)}, 798--808,
  \doi{10.1175/1520-0426(1993)010<0798:tddodv>2.0.co;2}.

\bibitem[{Kepner et~al.(2017)}]{kepner2017llgrid}
Kepner, J., and Coauthors, 2017: {LLGrid}: Supercomputer for sensor processing.
  \textit{Contemporary High Performance Computing}, Chapman and Hall/CRC,
  637--645.

\bibitem[{Kingma and Ba(2014)Kingma, and Ba}]{kingma2014adam}
Kingma, D.~P., and J.~Ba, 2014: Adam: A method for stochastic optimization.
  \textit{arXiv preprint arXiv:1412.6980}.

\bibitem[{Kollias et~al.(2022)}]{kollias++22}
Kollias, P., and Coauthors, 2022: {Science applications of phased array
  radars}. \textit{Bull. Amer. Meteor. Soc.}, in press,
  \doi{10.1175/bams-d-21-0173.1}.

\bibitem[{Krinitskiy et~al.(2018)Krinitskiy, Verezemskaya, Grashchenkov,
  Tilinina, Gulev,, and Lazzara}]{Krinitskiy2018DeepCN}
Krinitskiy, M., P.~Verezemskaya, K.~Grashchenkov, N.~Tilinina, S.~Gulev, and
  M.~Lazzara, 2018: Deep convolutional neural networks capabilities for binary
  classification of polar mesocyclones in satellite mosaics.
  \textit{Atmosphere}, \textbf{9~(11)}, 426.

\bibitem[{Kristovich et~al.(2003)Kristovich, Laird,, and
  Hjelmfelt}]{kristovich++03}
Kristovich, D. A.~R., N.~F. Laird, and M.~R. Hjelmfelt, 2003: {Convective
  evolution across Lake Michigan during a widespread lake-effect snow event}.
  \textit{Mon. Wea. Rev.}, \textbf{131~(4)}, 643--655,
  \doi{10.1175/1520-0493(2003)131<0643:cealmd>2.0.co;2}.

\bibitem[{Kurdzo et~al.(2017{\natexlab{a}})Kurdzo, Bennett, Veillette, Smalley,
  Williams,, and Donovan}]{kurdzo++17b}
Kurdzo, J.~M., B.~J. Bennett, M.~S. Veillette, D.~J. Smalley, E.~R. Williams,
  and M.~F. Donovan, 2017{\natexlab{a}}: {WSR-88D chaff detection and
  characterization using an optimized hydrometeor classification algorithm}.
  \textit{18th Conference on Aviation, Range, and Aerospace Meteorology}.

\bibitem[{Kurdzo et~al.(2015)Kurdzo, Bodine, Cheong,, and Palmer}]{kurdzo++15a}
Kurdzo, J.~M., D.~J. Bodine, B.~L. Cheong, and R.~D. Palmer, 2015:
  {High-temporal resolution polarimetric X-band Doppler radar observations of
  the 20 May 2013 Moore, Oklahoma, tornado}. \textit{Mon. Wea. Rev.},
  \textbf{143~(7)}, 2711--2735.

\bibitem[{Kurdzo et~al.(2017{\natexlab{b}})}]{kurdzo++17}
Kurdzo, J.~M., and Coauthors, 2017{\natexlab{b}}: {Observations of severe local
  storms and tornadoes with the atmospheric imaging radar}. \textit{Bull. Amer.
  Meteor. Soc.}, \textbf{98~(5)}, 915--935.

\bibitem[{Louf et~al.(2020)Louf, Protat, Jackson, Collis,, and
  Helmus}]{louf++20}
Louf, V., A.~Protat, R.~C. Jackson, S.~M. Collis, and J.~Helmus, 2020:
  {UNRAVEL: A robust modular velocity dealiasing technique for Doppler radar}.
  \textit{J. Atmos. Oceanic Tech.}, \textbf{37~(5)}, 741--758,
  \doi{10.1175/jtech-d-19-0020.1}.

\bibitem[{Michelson et~al.(1990)Michelson, Shrader,, and
  Wieler}]{michelson++90}
Michelson, M., W.~Shrader, and J.~Wieler, 1990: {Terminal Doppler weather
  radar}. \textit{Microwave Journal}, \textbf{33}, 139--148.

\bibitem[{Mitchell et~al.(1998)Mitchell, Vasiloff, Stumpf, Witt, Eilts,
  Johnson,, and Thomas}]{mitchell++98}
Mitchell, E. D.~W., S.~V. Vasiloff, G.~J. Stumpf, A.~Witt, M.~D. Eilts, J.~T.
  Johnson, and K.~W. Thomas, 1998: {The national severe storms laboratory
  tornado detection algorithm}. \textit{Wea. Forecasting.}, \textbf{13~(2)},
  352--366.

\bibitem[{Ortega et~al.(2016)Ortega, Krause,, and Ryzhkov}]{ortega++16}
Ortega, K.~L., J.~M. Krause, and A.~V. Ryzhkov, 2016: {Polarimetric radar
  characteristics of melting hail. Part III: Validation of the algorithm for
  hail size discrimination}. \textit{J. Appl. Meteor. Climatol.},
  \textbf{55~(4)}, 829--848, \doi{10.1175/jamc-d-15-0203.1}.

\bibitem[{Oye et~al.(1995)Oye, Mueller,, and Smith}]{oye++95}
Oye, R.~C., K.~Mueller, and S.~Smith, 1995: {Software for radar translation,
  editing, and interpolation}. \textit{27th Conf. on Radar Meteorology}.

\bibitem[{Palmer et~al.(2022)}]{palmer++22}
Palmer, R., and Coauthors, 2022: {A primer on phased array radar technology for
  the atmospheric sciences}. \textit{Bull. Amer. Meteor. Soc.},
  \doi{10.1175/bams-d-21-0172.1}.

\bibitem[{Park et~al.(2009)Park, Ryzhkov, Zrnic,, and Kim}]{park++09}
Park, H.~S., A.~V. Ryzhkov, D.~S. Zrnic, and K.-E. Kim, 2009: {The hydrometeor
  classification algorithm for the polarimetric WSR-88D: Description and
  application to an MCS}. \textit{Wea. Forecasting}, \textbf{24~(3)}, 730--748.

\bibitem[{Pazmany et~al.(2013)Pazmany, Mead, Bluestein, Snyder,, and
  Houser}]{pazmany++13}
Pazmany, A.~L., J.~B. Mead, H.~B. Bluestein, J.~C. Snyder, and J.~B. Houser,
  2013: {A mobile rapid-scanning X-band polarimetric (RaXPol) Doppler radar
  system}. \textit{J. Atmos. Oceanic Technol.}, \textbf{30~(7)}, 1398--1413.

\bibitem[{Peng et~al.(2022)Peng, Li,, and Jing}]{peng++21}
Peng, X., Q.~Li, and J.~Jing, 2022: {CNGAT: A graph neural network model for
  radar quantitative precipitation estimation}. \textit{IEEE Trans. on
  Geoscience and Remote Sensing}, \textbf{60}, 1--14,
  \doi{10.1109/tgrs.2021.3120218}.

\bibitem[{Racah et~al.(2017)Racah, Beckham, Maharaj, Kahou, Prabhat,, and
  Pal}]{Racah2017ExtremeWeatherAL}
Racah, E., C.~Beckham, T.~Maharaj, S.~E. Kahou, Prabhat, and C.~J. Pal, 2017:
  Extremeweather: A large-scale climate dataset for semi-supervised detection,
  localization, and understanding of extreme weather events. \textit{NIPS}.

\bibitem[{Ravuri et~al.(2021)}]{ravuri2021skilful}
Ravuri, S., and Coauthors, 2021: Skillful precipitation nowcasting using deep
  generative models of radar. \textit{Nature}, \textbf{597~(7878)}, 672--677.

\bibitem[{Ray and Ziegler(1977)Ray, and Ziegler}]{ray+77}
Ray, P.~S., and C.~Ziegler, 1977: {De-aliasing first-moment Doppler estimates}.
  \textit{J. Appl. Meteorol.}, \textbf{16~(5)}, 563--564,
  \doi{10.1175/1520-0450(1977)016<0563:dafmde>2.0.co;2}.

\bibitem[{Ronneberger et~al.(2015)Ronneberger, Fischer,, and
  Brox}]{ronneberger2015u}
Ronneberger, O., P.~Fischer, and T.~Brox, 2015: U-net: Convolutional networks
  for biomedical image segmentation. \textit{International Conference on
  Medical Image Computing and Computer-Assisted Intervention}, Springer,
  234--241.

\bibitem[{Samsi et~al.(2020)Samsi, Jones,, and Veillette}]{samsi2020compute}
Samsi, S., M.~Jones, and M.~M. Veillette, 2020: Compute, time and energy
  characterization of encoder-decoder networks with automatic mixed precision
  training. \textit{2020 IEEE High Performance Extreme Computing Conference
  (HPEC)}, IEEE, 1--6.

\bibitem[{Skolnik(2002)}]{skolnik02}
Skolnik, M., 2002: \textit{{Introduction to Radar Systems}}. 3rd ed.,
  McGraw-Hill.

\bibitem[{Stensrud et~al.(2009)}]{stensrud++09}
Stensrud, D.~J., and Coauthors, 2009: {Convective-scale warn-on-forecast
  system}. \textit{Bull. Amer. Meteor. Soc.}, \textbf{90}, 1487--1499.

\bibitem[{Stumpf et~al.(1998)Stumpf, Witt, Mitchell, Spencer, Johnson, Eilts,
  Thomas,, and Burgess}]{stumpf++98}
Stumpf, G.~J., A.~Witt, E.~D. Mitchell, P.~L. Spencer, J.~T. Johnson, M.~D.
  Eilts, K.~W. Thomas, and D.~W. Burgess, 1998: {The national severe storms
  laboratory mesocyclone detection algorithm for the WSR-88D}. \textit{Wea.
  Forecasting}, \textbf{13~(2)}, 304--326.

\bibitem[{Veillette et~al.(2018)Veillette, Hassey, Mattioli, Iskenderian,, and
  Lamey}]{veillette++18}
Veillette, M.~S., E.~P. Hassey, C.~J. Mattioli, H.~Iskenderian, and P.~M.
  Lamey, 2018: {Creating synthetic radar imagery using convolutional neural
  networks}. \textit{J. Atmos. Oceanic Technol.}, \textbf{35~(12)}, 2323 --
  2338.

\bibitem[{Wang et~al.(2021)Wang, Tang,, and Gentine}]{wang2021precipgan}
Wang, C., G.~Tang, and P.~Gentine, 2021: {PrecipGAN}: Merging microwave and
  infrared data for satellite precipitation estimation using generative
  adversarial network. \textit{Geophysical Research Letters}, \textbf{48~(5)},
  e2020GL092\,032.

\bibitem[{Weber et~al.(2021)}]{weber++21}
Weber, M., and Coauthors, 2021: {Towards the next generation operational
  meteorological radar}. \textit{Bull. Amer. Meteor. Soc.}, \textbf{102~(7)},
  E1357--E1383, \doi{10.1175/bams-d-20-0067.1}.

\bibitem[{Weyn et~al.(2021)Weyn, Durran, Caruana,, and
  Cresswell-Clay}]{weyn2021sub}
Weyn, J.~A., D.~R. Durran, R.~Caruana, and N.~Cresswell-Clay, 2021:
  Sub-seasonal forecasting with a large ensemble of deep-learning weather
  prediction models. \textit{Journal of Advances in Modeling Earth Systems},
  \textbf{13~(7)}, e2021MS002\,502.

\bibitem[{Witt et~al.(2009)Witt, Brown,, and Jing}]{witt++09}
Witt, A., R.~A. Brown, and Z.~Jing, 2009: {Performance of a new velocity
  dealiasing algorithm for the WSR-88D}. \textit{34th Conf. on Radar
  Meteorology}.

\bibitem[{Wurman et~al.(1996)Wurman, Straka,, and Rasmussen}]{wurman++96}
Wurman, J., J.~M. Straka, and E.~N. Rasmussen, 1996: {Fine-scale Doppler radar
  observations of tornadoes}. \textit{Science}, \textbf{272}, 1774--1777.

\bibitem[{Xing et~al.(2022)Xing, Hou, Huang,, and
  Zhang}]{xing2022spatiotemporal}
Xing, D., J.~Hou, C.~Huang, and W.~Zhang, 2022: Spatiotemporal reconstruction
  of {MODIS} normalized difference snow index products using {U-Net} with
  partial convolutions. \textit{Remote Sensing}, \textbf{14~(8)}, 1795.

\bibitem[{Xingjian et~al.(2015)Xingjian, Chen, Wang, Yeung, Wong,, and
  Woo}]{xingjian2015convolutional}
Xingjian, S., Z.~Chen, H.~Wang, D.-Y. Yeung, W.-K. Wong, and W.-c. Woo, 2015:
  Convolutional {LSTM} network: A machine learning approach for precipitation
  nowcasting. \textit{Advances in neural information processing systems},
  802--810.

\bibitem[{Xu et~al.(2011)Xu, Nai, Wei, Zhang, Liu,, and Parrish}]{xu++11}
Xu, Q., K.~Nai, L.~Wei, P.~Zhang, S.~Liu, and D.~Parrish, 2011: {A VAD-based
  dealiasing method for radar velocity data quality control}. \textit{J. Atmos.
  Oceanic Technol.}, \textbf{28~(1)}, 50--62, \doi{10.1175/2010jtecha1444.1}.

\bibitem[{Yang et~al.(2019)Yang, Zhao, Zhang, Chen, Huang,, and Chen}]{yang+19}
Yang, J., K.~Zhao, G.~Zhang, G.~Chen, H.~Huang, and H.~Chen, 2019: {A Bayesian
  hydrometeor classification algorithm for C-band polarimetric radar}.
  \textit{Remote Sensing}, \textbf{11~(16)}, 1884, \doi{10.3390/rs11161884}.

\bibitem[{Yo et~al.(2021)Yo, Su, Chu, Chang,, and Kuo}]{yo++21}
Yo, T., S.~Su, J.~Chu, C.~Chang, and H.~Kuo, 2021: {A deep learning approach to
  radar-based QPE}. \textit{Earth and Space Science}, \textbf{8~(3)},
  \doi{10.1029/2020ea001340}.

\bibitem[{Yu et~al.(2007)Yu, Wang, Shapiro, Yeary, Zrnic,, and
  Doviak}]{yu++07a}
Yu, T.-Y., Y.~Wang, A.~Shapiro, M.~B. Yeary, D.~S. Zrnic, and R.~J. Doviak,
  2007: {Characterization of tornado spectral signatures using higher-order
  spectra}. \textit{J. Atmos. Oceanic Tech.}, \textbf{24~(12)}, 1997--2013.

\bibitem[{Zhang and Wang(2006)Zhang, and Wang}]{zhang+06}
Zhang, J., and S.~Wang, 2006: {An automated 2D multipass Doppler radar velocity
  dealiasing dcheme}. \textit{J. Atmos. Oceanic Tech.}, \textbf{23~(9)},
  1239--1248, \doi{10.1175/jtech1910.1}.

\bibitem[{Zittel et~al.(2011)Zittel, Jing,, and Langlieb}]{zittel++11}
Zittel, W.~D., Z.~Jing, and N.~Langlieb, 2011: {A two-dimensional velocity
  dealiasing algorithm for the WSR-88D}. \textit{65th Interdepartmental
  Hurricane Conference}.

\end{thebibliography}

\end{document}